\documentclass[pre,amsmath,onecolumn, showpacs, superscriptaddress,10pt]{revtex4-1}
\usepackage{amssymb}
\usepackage{amsmath}
\usepackage{hyperref}
\usepackage{graphicx}

\usepackage{color}
\definecolor{Blue}{rgb}{0.00, 0.00, 1.00}
\definecolor{Red}{rgb}{1.00, 0.00, 0.00}

\newcommand{\bea}{\begin{eqnarray}}
\newcommand{\eea}{\end{eqnarray}}
\newcommand{\be}{\begin{equation}}
\newcommand{\ee}{\end{equation}}
\newcommand{\Ai}{{\rm Ai}}

\begin{document}

\title{Periodic Airy process and equilibrium dynamics of edge fermions in a trap}

\author{Pierre Le Doussal}
\affiliation{CNRS-Laboratoire de Physique Th\'eorique de l'Ecole Normale Sup\'erieure, 24 rue Lhomond, 75231 Paris Cedex, France}
\author{Satya N. \surname{Majumdar}}
\affiliation{LPTMS, CNRS, Univ. Paris-Sud, Universit\'e Paris-Saclay, 91405 Orsay, France}
\author{Gr\'egory \surname{Schehr}}
\affiliation{LPTMS, CNRS, Univ. Paris-Sud, Universit\'e Paris-Saclay, 91405 Orsay, France}

\begin{abstract} 
We establish an exact mapping between (i) the equilibrium (imaginary time) dynamics of non-interacting fermions trapped in a harmonic potential
at temperature $T=1/\beta$ and (ii) non-intersecting Ornstein-Uhlenbeck (OU) particles constrained 
to return to their initial positions after time $\beta$. Exploiting the determinantal structure of the process we compute the universal correlation functions both in the bulk and at the edge of the trapped Fermi gas.
The latter corresponds to the top path of the non-intersecting OU particles, and leads us to introduce and study the time-periodic Airy$_2$ process, ${\cal A}^b_2(u)$, depending on a single parameter, the period $b$. 
The standard Airy$_2$ process is recovered for $b=+\infty$. We discuss applications of our results to the real time quantum dynamics of trapped fermions. 

\end{abstract}


\maketitle

\tableofcontents

\section{Introduction}

\subsection{Background: from Ornstein-Uhlenbeck process to Airy$_2$ process}

The Airy$_2$ process was introduced in Ref.~\cite{PraSpo02} in the context of the discrete space, continuous time polynuclear growth model. Since then this process has appeared in many different contexts, such
as directed last passage percolation \cite{Joh03}, Dyson's Brownian motion \cite{Dys62}, non-intersecting Brownian bridges and excursions (watermelons) \cite{TW07,CH11}, growth models \cite{spohn_oup}, random tilings \cite{Joh05}, 
interacting particle transport \cite{Ferrari} and in the continuum KPZ equation \cite{CH16} (for a review see \cite{QR14,Baik_book}). To understand this process
one can consider the following simple example. Consider a single Ornstein-Uhlenbeck (OU) process \cite{Risken}
where the
position of a particle $x(\tau)$ in a one-dimensional harmonic potential evolves by the Langevin equation:
\be
\frac{d x(\tau)}{d \tau} = - \mu_0 \, x(\tau) + \eta(\tau) \;,  \label{OU1} 
\ee
where $\eta(\tau)$ is a centered Gaussian white noise, with 
correlator $\overline{ \eta(\tau) \eta(\tau')}=\delta(\tau-\tau')$. Consider the process starting at $x_0$ at time $\tau_0$.
The probability density $P_{\rm OU}(x,\tau|x_0,\tau_0)$ (the so called classical OU propagator) to arrive at $x$ at 
time $\tau \geq \tau_0$ has a Gaussian form 
\be
P_{\rm OU}(x,\tau|x_0,\tau_0) = \sqrt{ \frac{\mu_0}{\pi (1- e^{-2 \mu_0 (\tau-\tau_0)})}} \exp\left( - \frac{\mu_0 (x-x_0 e^{- \mu_0 (\tau-\tau_0)})^2}{1 - e^{-2 \mu_0 (\tau-\tau_0)}} \right)  \label{OUprop} \;.
\ee 
We now define an OU bridge over the time interval $[\tau_0,\tau_f]$ as a conditioned OU process that 
connects the initial position $x_0$ at $\tau_0$ and the final position $x_0$ at time $\tau_f$. To calculate the probability density of
the OU bridge at time $\tau \in [\tau_0,\tau]$, we split the time interval into two pieces: $[\tau_0, \tau]$ and $[\tau,\tau_f]$. For the first interval, the propagator is given in Eq. (\ref{OUprop}). For the second interval, we consider a further propagation of this process from time $\tau$ to time $\tau_f \geq \tau$ such that the final position is again $x_0$ at time $\tau_f$. Therefore, the probability density function (PDF) at $x$ at an intermediate time $\tau_0 \leq \tau \leq \tau_f$ of the OU bridge over the time interval $[0,\tau_f]$ is given by the product
\be\label{Pbridge}
P_{\rm bridge}(x,\tau) = \frac{P_{\rm OU}(x,\tau|x_0,\tau_0) P_{\rm OU}(x_0,\tau_f|x,\tau)}{P_{\rm OU}(x_0,\tau_f|x_0,\tau_0)} \;.
\ee  
The denominator in Eq. (\ref{Pbridge}) ensures that $P_{\rm bridge}(x,\tau)$ is normalized to unity, i.e., $\int_{-\infty}^\infty P_{\rm bridge}(x,\tau) \, dx = 1$, for any $\tau_0 \leq \tau \leq \tau_f$. If we now take the limits $\tau_0 \to -\infty$ and $\tau_f \to + \infty$, it is easy to see from Eqs. (\ref{OUprop}) and (\ref{Pbridge}) that $P_{\rm bridge}(x,\tau)$ is independent of $\tau$ and is given by 
\be
P_{\rm stat}(x) = \sqrt{\frac{\mu_0}{\pi}} \, e^{-\mu_0\,x^2} \label{Gibbs1} \;,
\ee
where the subscript ``stat'' indicates that this PDF is stationary, i.e., independent of time $\tau$. Note that $P_{\rm stat}(x)$ 
for the OU bridge in Eq. (\ref{Gibbs1}) is also the large $\tau$ limit of the standard OU propagator in Eq. (\ref{OUprop}) (identical to the equilibrium Gibbs measure of a particle in a quadratic potential $\mu_0 x^2/2$ at temperature $1/2$). In fact, this OU bridge process can also be identified to a time-periodic OU process (to be elaborated later in the paper) with period infinity~{\cite{footnote0}.

We can now generalize this OU bridge to an $N$-particle system, i.e, $N$ OU bridges, conditioned not to cross each other, $x_1(\tau) > x_2(\tau) > \cdots > x_N(\tau)$, at any time $\tau_0\leq \tau \leq \tau_f$. As in the case of $N=1$, the $N$-body free-propagator over the time interval $[\tau_0,\tau]$, such that they do not cross each other, can be obtained using the celebrated Karlin-McGregor formula \cite{KM59}. Similarly, one can write the propagator for the $N$ non-crossing OU paths over the right interval $[\tau,t]$. Then, as in the $N=1$ case (\ref{Pbridge}) , we take the product of the two propagators. We then take the limits $\tau_0 \to -\infty$ and $\tau_f \to + \infty$, assuming that they started close the origin (respecting the above ordering) at time $\tau_0 \to -\infty$ and are conditioned to return to the same initial positions at time $\tau_f \to + \infty$. As in the $N=1$ case, the joint PDF (JPDF) of the positions of the $N$ OU bridges become independent of $\tau$ and is given by~\cite{bray_winkler}
%
%
%
%assuming that  
%
%
%We assume that they started close to the origin (respecting the above ordering) at time $t_0 \to -\infty$ and are conditioned to return to the same initial positions at time $t \to + \infty$. 
%
%
%Again we can make the left-right decomposition of the paths at any intermediate time $\tau$, and take the product of the left and the right $N$-body 
%
%
%
%For the left part, the probability density that the $N$ walkers reach $x_1, x_2, \cdots, x_N$ at time $\tau$ again becomes independent of $\tau$ and is given by (see for instance \cite{Fisher84})
%\bea
%P_{\rm left}(x_1, \cdots, x_N, \tau)  \propto e^{- \frac{\mu_0}{2} \sum_{j=1}^N x_j^2}  \prod_{1\leq j< i \leq N} (x_i-x_j) \;. \label{LeftN}
%\eea 
%Incidentally this is also the joint PDF (JPDF) of the $N$ eigenvalues of an $N \times N$ real, symmetric Gaussian random matrix [known as the Gaussian Orthogonal Ensemble (GOE) \cite{meh91}]. Similarly, for the right part, considering the time reversed trajectories, we get
%\bea
%P_{\rm right}(x_1, \cdots, x_N, \tau)  \propto e^{- \frac{\mu_0}{2} \sum_{j=1}^N x_j^2}  \prod_{1\leq j < i \leq N} (x_i-x_j) \;. \label{RightN}
%\eea   
%Hence, taking the product, we obtain the JPDF of these non-intersecting OU bridges as
%%We assume that these processes started at an initial time $t_0 \to -\infty$
%%so that at any finite observation time $\tau$ they are already in the $N$-particle stationary state. It is
%%well known that in this state the joint PDF (JPDF) of all $\{ x_i(\tau) \}_{i=1,..N}$ at one given $\tau$ is given by %\cite{Dys62} 
\bea
P_{\rm stat}(x_1, \cdots, x_N) = A_N e^{- \mu_0 \sum_{j=1}^N x_j^2}  \prod_{j=1}^N (x_i-x_j)^2  \label{GibbsN} 
\eea 
for $x_1(\tau) > x_2(\tau) > \cdots > x_N(\tau)$, and zero otherwise, where $A_N$ is a normalization 
constant. This JPDF is isomorphic to the JPDF of the eigenvalues of a random matrix belonging to the
Gaussian Unitary Ensemble (GUE) \cite{meh91}. As in the $N=1$ case, these non-intersecting OU bridge processes can be identified to the non-crossing time-periodic OU paths with period infinity (see later).    

We now consider the trajectories of these non-intersecting OU bridges, {and study the correlations between observables at different times}. In particular let us consider the position of the top path $x_1(\tau)$ as a function of $\tau$.
This process $x_1(\tau)$, properly centered and scaled, converges in the large $N$ limit to a stationary
process, ${\cal A}_{2}(u)$, as a function of a rescaled time $u$ (to be defined more precisely later), called the Airy$_2$ process. 
%(in units such that $\mu_0=1$ {\red I dont understand that condition, I would remove it it is
%confusing, or am I missing something? you already said that it is rescaled etc..}). 
This construction of the Airy$_2$ process presented here is equivalent to the similar construction using Dyson's Brownian motion~\cite{PraSpo02,Baik_book}. This particular construction is useful for us for generalization to finite temperature later. 

Note that in a system of non-intersecting Brownian bridges (known as the watermelon problem \cite{PraSpo02, Joh03, TW07, watermelon_us,KIK08,FMS11,SMCF13}) the statistics of the top path, properly centered and scaled, is described by ${\cal A}_2(u) - u^2$, where $u$ is again the rescaled time. Likewise, in the context of the (1+1)-dimensional KPZ class growth models in the curved (droplet) geometry, the {rescaled} height $h(x,t)$ of the interface, at a fixed large time $t$, converges to ${\cal A}_2(u)-u^2$ (where $u$ in this case the represents rescaled spatial
position $x$)~\cite{PraSpo02, Joh03, Cor12,KK10}.} 

Let us now describe briefly the main characteristics of the Airy$_2$ process. Its one point marginal distribution at fixed time $\tau$ is given by the Tracy-Widom GUE distribution $F_2(s)$  \cite{TW94}
\be
{\rm Prob}( {\cal A}_{2}(u) < s ) = F_2(s) = {\rm Det}[ I - P_s K_{\Ai} P_s] = \exp\left( {\rm Tr} \ln(  I - P_s K_{\Ai} P_s) \right) \label{Airy1pt0} 
\ee
where $I$ is the identity, $P_s$ is the projector on the interval $[s,+\infty[$ and ${\rm Det}$ stands for a 
Fredholm determinant on the space $L^2(\mathbb{R})$ with the Airy kernel
\be
K_{\Ai}(r,r') = \int_0^{+\infty} dv \, \Ai(r+v) \Ai(r'+v) \;.
\ee
Note that Eq. \eqref{Airy1pt0} can be evaluated in several ways, one being to use the expansion ${\rm Tr} \ln (I - A) =
- \sum_{p=1}^{+\infty} \frac{1}{p} {\rm Tr} A^p$.

In addition, one can generalize this one time distribution to multi-time JPDF. 
For example the 
two-time joint cumulative distribution function (JCDF) reads \cite{PraSpo02,QR14}
\be
{\rm Prob}( {\cal A}_{2}(u_1) < s_1, {\cal A}_{2}(u_2) < s_2) 
= {\rm Det}
\begin{pmatrix} I - P_{s_1} K_{u_1,u_1} P_{s_1} & -  P_{s_1} K_{u_1,u_2} P_{s_2}   \\  
 - P_{s_2} K_{u_2,u_1} P_{s_1}  & I - P_{s_2} K_{u_2,u_2} P_{s_2}  \end{pmatrix} \;,
 \label{Airy2pt0} 
\ee 
where $K_{u,u'}$ is the {\it extended Airy kernel} defined as
\bea \label{AiryExtended} 
K_{u,u'}(r,r') =
\begin{cases}
&  ~ \int_0^{+\infty} dv \, \Ai(r+v) \Ai(r'+v) e^{- v(u-u')} , \quad {\rm for} \quad u \geq u' \\
&\\
& -  \int_{-\infty}^0 dv \, \Ai(r+v) \Ai(r'+v) e^{- v(u-u')} , \quad {\rm for} \quad u < u'  \;.
\end{cases}
\eea
The right hand side of Eq. \eqref{Airy2pt0} is a generalization of the standard Fredholm determinant, where the
kernel itself has a $2 \times 2$ matrix structure. To evaluate the Fredholm determinant in
\eqref{Airy2pt0} one may use, as above, an expansion in traces of powers of the matrix-operator in 
\eqref{Airy2pt0}, which has a block structure (for an example
see section \ref{sec:applications}). {By analyzing the asymptotics of this extended kernel, one can show that
the connected part of the correlation function is stationary and has a power law tail \cite{PraSpo02,AM04}
\bea \label{asympt_correl}
{\rm Prob}( {\cal A}_{2}(u_1) < s_1, {\cal A}_{2}(u_2) < s_2) - {\rm Prob}( {\cal A}_{2}(u_1) < s_1){\rm Prob}( {\cal A}_{2}(u_2) < s_2) \sim \frac{1}{|u_1-u_2|^2} \;, 
\eea
for large $|u_1 - u_2|$.} Similarly the $n$-time correlation
function of the Airy$_2$ process can be written as a Fredholm determinant, where the
kernel itself has a $n \times n$ matrix structure. This property is a consequence of an underlying determinantal structure which will be explained later. 

Another interesting application of this process, which we will develop extensively in this paper, concerns
the quantum mechanical problem of $N$ non-interacting fermions in a harmonic trap at zero temperature. In a recent series of papers we have studied the connection between the position of the trapped fermions $x_j$
and the eigenvalues of the GUE random matrix \cite{MMSV14,MMSV16,DLMS15,dea15b,DLMS16,dea16} (see also \cite{Eis13,Castillo14}). Indeed the quantum JPDF for the positions of the fermions in the many-body ground state is identical to the JPDF of the GUE eigenvalues, given 
in \eqref{GibbsN}. As a consequence, the average density of fermions in the trap at
zero temperature, $T=0$, in the large $N$ limit, is given by the Wigner semi-circular law for the GUE eigenvalues
which vanishes outside a finite interval. Hence this predicts that the Fermi gas in a
harmonic trap has a sharp edge where the density vanishes. 
One immediate consequence of this mapping is that the
position of the rightmost fermion, $x_{\max}$ at $T=0$, suitably centered and scaled,
converges in the large $N$ limit, to the GUE Tracy-Widom distribution (\ref{Airy1pt0}). In addition the suitably scaled $n$-point correlation functions of 
the Fermi gas near the edge, can be expressed as an $n \times n$ determinant
based on the Airy kernel. In \cite{Eis13} these properties have been shown to hold more generally
for other smooth trapping potentials. 

Moreover, these zero temperature properties of the trapped Fermi gas were extended
to finite temperature $T>0$~\cite{DLMS15,DLMS16}. In the large $N$ limit, the determinantal structure of the correlation functions persist
even at finite temperature, with a $T$-dependent kernel, generalizing the $T=0$ Airy kernel. 
Interestingly the same kernel was found to appear in the solution of the continuum 1+1 dimensional KPZ equation at finite time \cite{DLMS15,DLMS16}.

\subsection{Summary of the main results of this paper}

All the above results on trapped fermions concern only static quantities at equilibrium. 
It is natural to ask how these systems of fermions at equilibrium evolve in imaginary time $\tau$ (which
can be seen as the analytic continuation of the real time equilibrium quantum dynamics of the fermions
$\tau = i t$). Similar questions have been studied in the condensed matter literature 
\cite{FW71,GiamBook,Pereira,Xavier,Sims,Stolze} but mostly in the bulk of the Fermi gas, or in the absence of
a confining potential. Our main focus here is on the dynamics at the edge, for which it is useful to exploit connections to 
 random matrices \cite{Mac94} and determinantal processes \cite{Borodin1}. 
In this paper we address the question of the dynamics of trapped non interacting fermions in both imaginary and real time. It is useful first to summarize our main results.
\begin{itemize}

\item First we establish an exact mapping between the quantum propagator in imaginary time of $N$ non interacting fermions in a harmonic potential, and the transition probability of a collection of $N$ non-intersecting classical Ornstein-Uhlenbeck (OU) processes. 

\item Next, we extend this mapping to the imaginary time dynamics of $N$ non interacting fermions in a harmonic potential at finite temperature $T=1/\beta$ and a collection of $N$ non-intersecting classical Ornstein-Uhlenbeck (OU) processes, periodic in time with period $\beta$ (see Fig. \ref{fig:cylinder}). 
In particular the positions of the fermions at the edge of the trapped Fermi gas correspond to the positions of the top paths of the 
collection of the time periodic OU processes. An immediate consequence of our results is that at $T=0$ for the fermions in the harmonic trap, the imaginary time dynamics of the rightmost fermion, properly centered and scaled, is the Airy$_2$ process, discussed in the introduction.
This is thus a new application of the Airy$_2$ process. Furthermore we show a direct mapping between the (imaginary) time-dependent fermion positions and the time-dependent eigenvalues in the Dyson's Brownian motion. 

\item
Using the Eynard-Mehta (EM) theorem \cite{EM98} known in random matrix theory, 
we show that both the fermion and the OU problems possess an extended determinantal structure, which we exploit to compute several correlation functions as determinants of an extended kernel. We provide, both in the bulk (\ref{extendedsine})-(\ref{extendedsine3}) of the Fermi gas and at the edge (\ref{scaledge})-(\ref{TextAiry1}), explicit forms both in space and time of the kernel at finite temperature in the large $N$ limit. We show that, for a large class of confining potentials, these scaling forms are universal, i.e. independent of the details of the trapping potential.
At the edge this universal kernel ${\cal K}_b^{\rm edge}$ depends on a single parameter, the
reduced inverse temperature $b= \mu_0 N^{1/3} \beta$.  

\item Recalling that the imaginary time dynamics of the position of rightmost fermion, $x_{\max}(\tau)$, in a harmonic potential at {\it zero temperature}, suitably scaled, converges in the large $N$ limit to the Airy$_2$ process, it is then natural to ask what is the process
$x_{\max}(\tau)$ {\it at finite temperature} $T=1/\beta>0$ in the large $N$ limit. Here we provide this generalization, which we call {\it the periodic Airy$_2$ process}. We denote this process
by ${\cal A}^b(u)$ where $u= b \tau/\beta$ is the rescaled time. This process is stationary, periodic of period $b$, and invariant in distribution under parity transformation $u \to -u$. We show that it also has
a determinantal structure which involves the universal kernel ${\cal K}_b^{\rm edge}$ given in Eqs. (\ref{scaledge})-(\ref{TextAiry0}).

\item The real time dynamics of the fermions both in the bulk and at the edge, at finite
temperature, can be described using an analytic continuation from our results in imaginary time.
In particular we provide the exact result for the two time density-density correlation function: while the result in the bulk can be related to results for free fermions known in the literature (with no trap), the results at the edge are non-trivial.
We show that the large time decay at the edge is $\sim t^{-3/2}$ at $T=0$, with a crossover to exponential decay at finite temperature. Finally, we also show that successive quantum measurements of the fermion positions can be addressed
by similar techniques.

\end{itemize}

The rest of the paper is organized as follows. In section \ref{sec:sec2}, as a warm up, we provide an exact correspondence 
between imaginary time quantum propagator of $N$ non interacting fermions in a harmonic potential and the
transition probability of a collection of $N$ non-intersecting classical Ornstein-Uhlenbeck (OU) processes.
In section \ref{sec:Nfermions} we extend this connection to finite temperature $T$ for the fermion problem, 
which then corresponds to non-intersecting classical OU paths periodic in time of period $\beta=1/T$.
In section \ref{sec:EM} we recall the Eynard-Mehta (EM) theorem and the associated determinantal structure for the multi-time (imaginary time) correlation function of the fermions. In section \ref{sec:ground} we apply the EM theorem to the ground state of fermions, and provide the scaling forms of the extended kernels in the bulk and at the edge at zero temperature.
In section \ref{sec:finiteT} we extend these results to finite temperature. In section \ref{sec:TAiry} we infer the corresponding consequences for the time-periodic OU processes. In particular, by considering the time evolution of the top path of the OU processes, we define and characterize the periodic Airy$_2$ process, ${\cal A}^b(u)$, where $b$ is the period, and calculate explicitly some asymptotic properties of the two-time correlation function of this process.
In section \ref{sec:univ} we consider non-intersecting time-periodic Brownian motions in more general potentials and argue that the top line is always described by the universal periodic Airy$_2$ process. In section \ref{sec:real}
we apply the results of the previous sections to study the real time quantum equilibrium dynamics of the fermions at the edge of the Fermi gas. Finally we conclude in section \ref{sec:conclusion}.

More details are provided in the Appendices. 
The Appendix \ref{app:DBM} contains an explicit mapping between the Dyson's Brownian motion, the non-crossing OU processes, and the fermions in an harmonic trap. The Appendix \ref{app:more} contains an explicit construction of a time-periodic
OU process. In Appendices \ref{app:details}, \ref{app:repro}
and \ref{app:det}, we provide more details on the Eynard-Mehta theorem, the extended
kernel and the multi-time correlation functions, respectively.

\section{Dynamics of $N$ fermions in a harmonic trap and $N$ non-crossing Ornstein-Uhlenbeck process} 
\label{sec:sec2} 

In this section we provide an exact mapping between the imaginary time quantum propagator of $N$ non interacting fermions in a harmonic trap and the transition probability of a collection of $N$ non-intersecting classical Ornstein-Uhlenbeck (OU) processes. We first recall the $N=1$ single particle case, and later provide the many-body generalization.

\subsection{Single particle} \label{sec:single} 

We start with the single particle OU process in Eq. \eqref{OU1}. The PDF $P(x,\tau)$ of the particle position 
satisfies the Fokker-Planck equation
\be
\partial_\tau P = \frac{1}{2} \, \partial^2_x P + \mu_0 \, \partial_x (x P) \;,
\ee 
with initial condition $P(x,\tau_0)=\delta(x-x_0)$ at time $\tau=\tau_0$. The solution is the OU propagator given in Eq. (\ref{OUprop})
%which reads
%\be
%P_{\rm OU}(x,\tau|x_0,\tau_0) = \sqrt{ \frac{\mu_0}{\pi (1- e^{-2 \mu_0 (\tau-\tau_0)})}} \exp\left( - \frac{\mu_0 (x-x_0 e^{- \mu_0 (\tau-\tau_0)})^2}{1 - e^{-2 \mu_0 (\tau-\tau_0)}} \right)  \label{OUprop} \;.
%\ee 
For $\tau_0 \to -\infty$ the distribution is time independent at any finite $\tau$ and is given by \eqref{Gibbs1}. 

We now recall how this propagator can be related to the quantum propagator in imaginary time of a simple harmonic oscillator with Hamiltonian
\be \label{osc}
\hat H = - \frac{\hbar^2}{2 m} \frac{\partial^2}{\partial x^2} + \frac{1}{2} m \omega^2 x^2 - \frac{\hbar \omega}{2} \;.
\ee 
A shift in the potential energy has been added for later convenience so that the quantized
energy levels are given by $\epsilon_k= \hbar \omega\, k$, where $k=0,1,2, \ldots$. 
The corresponding eigenfunctions are denoted by $\phi_k(x)$. 
From now on, for simplicity, we will work in the units where $\hbar=m=1$ and set $\omega=\mu_0$
to make the correspondence between the OU process and the quantum problem.
In these units the normalized eigenfunctions are
\be
\phi_k(x) = \left(\frac{\sqrt{\omega}}{\sqrt{\pi} 2^k k!}\right)^{1/2} e^{- \frac{\omega \,x^2}{2}} H_k(\sqrt{\omega} x) \;,
\label{Hermite} 
\ee
where $H_k$ is the Hermite polynomial of degree $k$. 
The imaginary time propagator for this harmonic oscillator, between $\tau_0$ and $\tau$
can be written in quantum mechanical notations as {\cite{footnotenew}}
\be \label{G} 
G(x,\tau|x_0, \tau_0) = \langle x| e^{- (\tau-\tau_0) \hat H} |x_0 \rangle 
= \sum_{k=0}^{+\infty} \phi_k(x)  \phi_k^*(x_0) e^{- \epsilon_k (\tau-\tau_0)} 
\ee 
and satifies 
\be
\partial_\tau G = - \hat H G  \;,
\ee 
with initial condition $G(x,\tau_0|x_0 \tau_0) = \delta(x-x_0)$. It is easy to check that
\bea \label{correspondence1}
P_{\rm OU}(x, \tau|x_0, \tau_0) = e^{-\mu_0 \frac{x^2}{2} } G(x,\tau|x_0, \tau_0)  \, e^{\mu_0 \frac{x_0^2}{2} } \;,
\label{POU3} 
\eea 
which identifies the quantum propagator in imaginary time of a single particle in a harmonic trap, with the
classical propagator of the OU process. 

\subsection{$N$ particles} \label{sec:N} 

Let us now start with the fermion problem. Consider $N$ non-interacting fermions in a one-dimensional harmonic trap 
with $N$-body Hamiltonian 
\be
{\cal H}_N = \sum_{j=1}^N \hat H_j \label{HN} \;,
\ee 
where the single particle Hamiltonian is given in \eqref{osc}. 
For non-interacting fermions each many-body eigenstate of ${\cal H}_N$ associated to energy $E$, and denoted by $\psi_E(x_1,\cdots,x_N)$, is a Slater determinant, normalized to unity, consisting of occupied single-particle states 
\be
\psi_E(x_1,\cdots,x_N) = \frac{1}{\sqrt{N!}} \det_{1\leq i,j \leq N} \phi_{k_i}(x_j)   \label{slater} \;,
\ee
where $k_1<k_2 <\cdots < k_N$ are the labels of the single particle eigenfunctions $\phi_{k}(x)$.
The quantum propagator in imaginary time of the $N$ fermion system can be written as
\be
G^{(N)}(x_1,\cdots,x_N;\tau| y_1,\cdots,y_N;0) =
\langle x_1,\cdots , x_N | e^{- {\cal H}_N \tau} | y_1, \cdots , y_N \rangle
=  \sum_E \psi_E(x_1,\cdots , x_N)
\psi_E^*(y_1, \cdots , y_N) e^{- E \tau}  \label{GN}
\ee
where the loose notation of summation over $E$ includes possible degeneracies,
and where the $\psi_E$'s are given in \eqref{slater}. Using the determinantal form
of the $\psi_E$'s in \eqref{slater} and the Cauchy-Binet formula for a discrete integration measure, we obtain a determinantal formula for the $N$ fermion quantum propagator
\bea
G^{(N)}(x_1,\cdots,x_N;\tau| y_1,\cdots,y_N;0) &=& 
\frac{1}{N!}  \sum_{0\leq k_1<k_2<\cdots <k_N} \det_{1\leq i,j \leq N} \phi_{k_i}(x_j)  \det_{1\leq i,j \leq N} \phi^*_{k_i}(y_j) e^{- \tau \sum_{\ell=1}^N \epsilon_{k_\ell} } 
\label{Satya} \\
&=& \frac{1}{N!^2}  \sum_{k_1,k_2,\cdots ,k_N=0}^{+\infty} 
 \det_{1\leq i,j \leq N} \phi_{k_i}(x_j)  \det_{1\leq i,j \leq N} \phi^*_{k_i}(y_j) e^{- \tau \sum_{\ell=1}^N \epsilon_{k_\ell} }  \\
 &=& \frac{1}{N!} \det_{1\leq i,j \leq N} G(x_i,\tau|y_j,0) \label{detG} \;,
\eea 
where $G(x_i,\tau|y_j,0)$ is the single particle quantum propagator given in equation \eqref{G}. Evidently, for $N=1$ it reduces to the single particle quantum propagator.

We now define the
transition probability 
$P^{(N)}_{\rm OU}(x_1,\cdots,x_N;\tau| y_1,\cdots,y_N;0)$ that a set of
$N$ distinguishable OU processes $x_i(t)$, $i=1,..N$, $0 \leq t \leq \tau$, starting at the initial positions 
$y_1>y_2>\cdots >y_N$ at time $0$ arrive at $x_1 > x_2 > \cdots >x_N$ 
at time $\tau$ {\it and} have not crossed each other in the time interval $t \in [0,\tau]$. For $N$ particles the
generalization of Eq. \eqref{POU3} can be written (in the ordered sector)
\bea
P^{(N)}_{\rm OU}(x_1,\cdots,x_N;\tau| y_1,\cdots,y_N;0) & = & N! \, e^{- \frac{\mu_0}{2} \sum_{i=1}^N x_i^2}
\, G^{(N)}(x_1,\cdots,x_N;\tau| y_1,\cdots,y_N;0)  \, e^{ \frac{\mu_0}{2} \sum_{i=1}^N y_i^2} \label{PN1} \;.
\eea
Note that the factor $N!$ comes from the fact that the $N$ OU processes are distinguishable, while the
corresponding quantum particles are indistinguishable. 
Using the
determinantal form \eqref{detG} for the quantum propagator, and Eq. \eqref{POU3} this transition probability can also be written as
\be
P^{(N)}_{\rm OU}(x_1,\cdots,x_N;\tau| y_1,\cdots,y_N;0) = \det_{1\leq i,j \leq N} P_{\rm OU}(x_i , \tau | y_j, 0) \, ,  \label{propOU2}
\ee
thus recovering the celebrated Karlin-McGregor formula for non intersecting paths \cite{KM59}, in this
particular case of OU processes. Note that Eqs. \eqref{PN1}, \eqref{propOU2} are valid only in the ordered sector of coordinates,
and furthermore that the transition probability $P^{(N)}_{\rm OU}$, when integrated over the ordered sector of $x_i$
coordinates, gives the probability that the $N$ paths do not cross each other up to time $\tau$. 

There also exist an interesting connection between the $N$ non-crossing OU paths 
and the Dyson's Brownian motion, which is detailed in Appendix \ref{app:DBM}. Using this correspondence
one can infer that the (imaginary time) evolution of the positions of the fermions maps onto 
the time evolution (via the DBM) of the eigenvalues of a GUE matrix. 

\section{$N$ fermions in a harmonic trap at finite temperature and non-crossing time-periodic OU paths}
\label{sec:Nfermions} 

So far we have been addressing the quantum-classical correspondence between the imaginary time dynamics
of N trapped fermions and the stochastic dynamics of N non-crossing OU processes. In this section, we consider
the single time properties of $N$ fermions prepared at thermal equilibrium at temperature $T$. We show that there is again a mapping to a classical stochastic process of $N$ non crossing OU paths, but the paths are now periodic in the time direction with period $\beta=1/T$. We call this process the time periodic non-crossing OU process. In the following sections
we will combine these two aspects (state preparation and dynamics) to study the multi-time correlations at thermal equilibrium. In this section also, we start with a single particle and then generalize to $N$ particles. 

\subsection{Single particle}

Consider a single quantum harmonic oscillator at finite temperature $T=1/\beta$ in the canonical ensemble. The PDF of the
position of the particle, obtained from the quantum density matrix, is 
\be \label{PTx} 
P_T(x) = \frac1{Z_1} \sum_{k=0}^{+\infty} |\phi_k(x)|^2 e^{- \beta \epsilon_k}  = \frac{G(x,\beta|x,0)}{Z_1}
\ee 
where $Z_1= \sum_{k=0}^{+\infty} e^{- \beta \epsilon_k}$ is the partition sum and the propagator $G$ is defined in Eq. (\ref{G}). 

\begin{figure}
\begin{center}
\includegraphics[width = 0.7\linewidth]{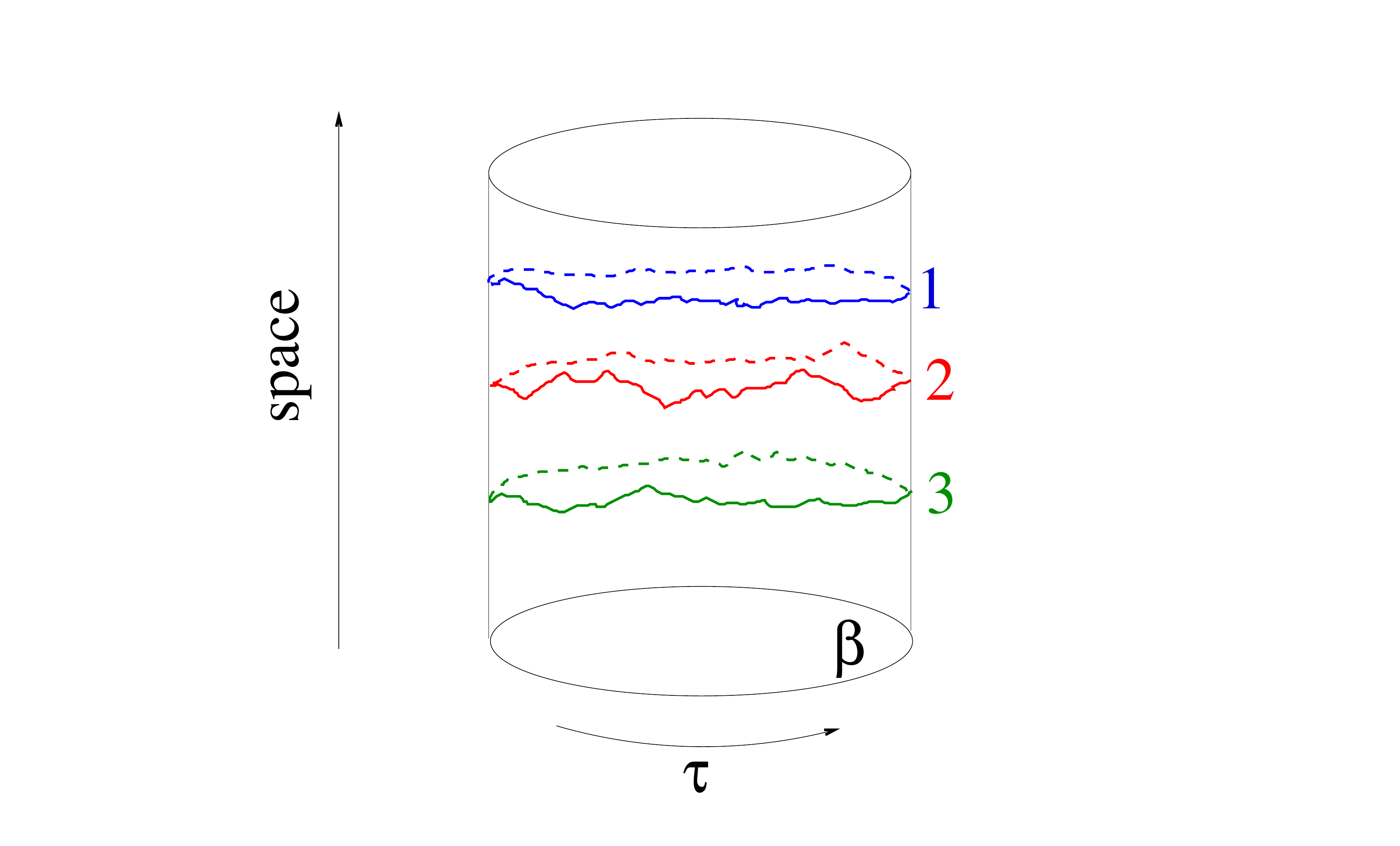}
\caption{$N=3$ nonintersecting OU processes wrapped around the cylinder
of perimeter $\beta$. The imaginary time $\tau$ runs anticlockwise
over the interval $\tau\in [0,\beta]$ where $\beta$ is the inverse
temperature. The vertical direction denotes space with 
$x_1(\tau)>x_2(\tau)>x_3(\tau)$ being the coordinates of the
three processes at time $\tau$.}
\label{fig:cylinder}
\end{center}  
\end{figure}

One can now ask how to interpret this quantity in terms of the OU process. Consider 
the OU process on the time interval $[0,\beta]$ with the condition that $x(0)=x(\beta)$. 
This is called the time periodic OU process, to emphasize that the periodicity here is in the time direction
and not in the spatial direction. Setting $x=x_0$ in \eqref{correspondence1}, it follows that
\be \label{PTx2} 
P_T(x) = \frac{1}{Z_1} P_{\rm OU}(x,\beta|x,0) \;,
\ee
where $P_{\rm OU}(x,\tau|x_0,0)$ is given in \eqref{OUprop}.
One can visualize this process as a fluctuating directed line 
wrapped around the cylinder of perimeter $\beta$ in a quadratic well, see Fig. 
\ref{fig:cylinder}. From Eq. \eqref{OUprop}, and evaluating the partition function $Z_1$, we obtain
\bea
P_T(x) = \sqrt{ \frac{\mu_0}{\pi} \tanh\left(\frac{\beta \mu_0}{2}\right) } \,\,
e^{ - \mu_0 \tanh\left(\frac{\beta \mu_0}{2}\right) x^2 } \label{statbeta}  \;.
\eea
One can check that $P_T(x)$ is normalized to unity upon integration over $x$. 
Note that although it is a simple Gaussian distribution, it actually involves a sum over all
excited states of the harmonic oscillator. Clearly, as $\beta \to +\infty$, i.e. $T \to 0$, one
obtains $P_0(x)=P_{\rm stat}(x)$ as in \eqref{Gibbs1}. Thus $P_T(x)$ in \eqref{statbeta}
is the stationary measure of the time-periodic OU process on a cylinder of perimeter $\beta$
(see Appendix \ref{app:more} for a more precise description).

\subsection{$N$ particles}

We consider the canonical ensemble with a finite temperature $T=1/\beta$.
The quantum JPDF of the $N$ fermion positions is given by the generalization
of the single particle formula in \eqref{PTx} 
\bea
P_T(x_1,\cdots,x_N) = \frac{1}{Z_N(\beta)} \sum_{E} |\psi_E(x_1,\cdots,x_N)|^2 e^{- \beta E} \label{PTxN} 
\eea 
where $\psi_E(x_1,\cdots,x_N)$ is the many-body energy eigenfunction with energy $E$
and the sum is over a complete orthonormal basis of eigenstates. The partition sum $Z_N(\beta)= \sum_{E} e^{- \beta E}$ is the normalization constant which ensures that $P_T$ is normalized to unity in $\mathbb{R}^N$. 
Note that $P_T$ is a symmetric function in all of its arguments. 

Hence one 
can rewrite the JPDF \eqref{PTxN} as
\bea
P_T(x_1,\cdots,x_N) = \frac{1}{N! Z_N(\beta)} \sum_{0 \leq k_1<\cdots < k_N} |\det_{1\leq i,j \leq N} \phi_{k_i}(x_j)|^2 e^{- \beta \sum_{i=1}^N \epsilon_{k_i}} \label{PTxN2} \;,
\eea
where $\epsilon_{k_i}$ are the eigenenergies of the occupied levels, and the sum is over all
possible fillings of the single particle levels by the fermions, with at most one fermion in a given level. 
The partition sum is thus 
\be
Z_N(\beta)=\sum_{E} e^{- \beta E} =\sum_{0\leq k_1<\cdots < k_N} e^{- \beta \sum_{i=1}^N \epsilon_{k_i}} \label{Zbeta} 
\ee
 and
one can explicitly check the normalization to unity of $P_T$ on $\mathbb{R}^N$.
Since the determinant vanishes for coinciding $k_i$ one can replace (using the invariance
under permutation of the $k_i$'s) the sum $\sum_{k_1<\cdots < k_N}$ by the unconstrained sum $\frac{1}{N!} \sum_{k_1,\cdots,k_N}$.
Next, using the discrete version of the Cauchy-Binet formula, one obtains
\bea
P_T(x_1,\cdots,x_N) = \frac{1}{N! Z_N(\beta)}  \det_{1\leq i,j \leq N} G(x_i , \beta | x_j, 0)  \label{PTxNLast} 
\eea 
where 
\bea
G(x, \beta | x', 0)  = \sum_{k=0}^{+\infty} \phi_k(x) \phi^*_k(x') e^{- \beta \epsilon_k} 
= \left(\frac{\omega}{2 \pi \sinh(\beta \omega)}\right)^{1/2}
\exp\left( - \frac{\omega}{2 \sinh(\beta \omega)} ( (x^2 + x'^2) \cosh(\beta \omega) - 2 x x') \right)
 \label{propdec}
\eea 
is just the single-particle harmonic oscillator propagator in imaginary time $\beta$. Note that $P_T(x_1,\cdots,x_N)$ is a symmetric and positive function
which integrates to unity on $\mathbb{R}^N$. Clearly for $N=1$ the formula \eqref{PTxNLast} for the JPDF
reduces to Eq. \eqref{PTx}. Using \eqref{correspondence1} we can rewrite equation
\eqref{PTxNLast} as 
\bea
P_T(x_1,\cdots,x_N) = \frac{1}{N! Z_N(\beta)}  \det_{1\leq i,j \leq N} P_{\rm OU}(x_i , \beta | x_j, 0)  \label{PTxN2} 
\eea 
%This determinant has an interpretation in terms of the Karlin-MacGregor formula for
%non-crossing paths. According to this formula the probability 
%$P^{(N)}_{\rm OU}(x_1,\cdots,x_N;\beta| x_1,\cdots,x_N;0)$ that a set of
%$N$ distinguishable OU processes $x_i(\tau)$, $i=1,..N$, $0 \leq \tau \leq \beta$, starting at the initial positions 
%$x_1,\cdots ,x_N$ at time $0$ arrive at the same positions $x_1,\cdots, x_N$
%at time $\tau$ {\it and} have not crossed each other in the time interval $[0,\beta]$
%is given by a determinant 
%\bea
%P^{(N)}_{\rm OU}(x_1,\cdots,x_N;\beta| x_1,\cdots,x_N,0) =  \det_{1\leq i,j \leq N} P_{\rm OU}(x_i , \beta | x_j, 0) \label{propOU}
%\eea 
Therefore we have from \eqref{propOU2}
\bea
P_T(x_1,\cdots,x_N) = \frac{1}{N! Z_N(\beta)}  
P^{(N)}_{\rm OU}(x_1,\cdots,x_N;\beta| x_1,\cdots,x_N,0)  \;. \label{PTxN3} 
\eea 
Note that since the arrival and final points here are the same, this identity is correct
for arbitrary ordering of the $x_i$. One can check that $P_T$ given by this formula is
normalized to unity. 
This is the generalization for $N$ particles, of the formula \eqref{PTx2} for 
$N=1$. We can visualize this as a set of $N$ distinguishable non-crossing OU paths wrapped
on the cylinder of perimeter $\beta$ as shown in Fig. \ref{fig:cylinder}.
%Integrating \eqref{PTxN3} on the sector $x_1>\cdots > x_N$, the l.h.s
%gives $1/N!$, hence the total probability of non-crossing
%\bea
%\int_{x_1>\cdots > x_N} P^{(N)}_{\rm OU}(x_1,\cdots,x_N;\beta| x_1,\cdots,x_N,0) = Z_N 
%\eea
%identifies with the fermion partition function.
This completes the correspondence between the quantum (equal time) JPDF 
 for the fermion problem, and a set of $N$ non-crossing time-periodic OU processes.
 A similar construction was used in the context of fluctuating $(1+1)$-dimensional non-intersecting interfaces
 in an external potential \cite{nad09}. Note that in the limit $T \to 0$ one can see that the JPDF $P_0(x_1,\cdots,x_N)$ becomes equal 
to the JPDF of the eigenvalues of a GUE random matrix as given in \eqref{GibbsN}. Hence
$P_T(x_1,\cdots,x_N)$ is the finite temperature generalization of this JPDF, but evaluated
at any fixed time $\tau \in [0,\beta]$.

\section{Eynard-Mehta theorem, non-interacting fermions and determinantal process} 
\label{sec:EM} 

In section \ref{sec:sec2} we have obtained the complete quantum propagator in imaginary time
for $N$ non interacting fermions. It has the form of a $N \times N$ determinant of single particle
propagators (\ref{detG}). It is however a daunting task to extract, from this large determinant, explicit results for
multi-time correlation functions. It turns out, however that if one prepares the system in 
any eigenstate $|E \rangle$ of the $N$ body Hamiltonian, it is possible to obtain more explicit
formulae for correlation function. In this section we demonstrate this fact. 
%
%The aim of this section is twofold. First we want to study the multi-time JPDF
%of both systems, i.e. the JPDF of the extended space-time process. 
We will show that the multi-time JPDF possess an extended
determinantal structure which allows to express any marginal JPDF (obtained by
integrating over any subset of positions) as a determinant
constructed from a single extended kernel. We show that this can
be achieved by exploiting the Eynard-Mehta (EM) theorem~\cite{EM98,BorodinOlshanski,TW07,Johansson2005}, which is
recalled below. For concreteness, we consider $N$ fermions in a harmonic trap, though the results of this section are actually valid for more general potentials.

\subsection{Joint multi-time PDF in a given many-body fermionic eigenstate} \label{sec:state}

To proceed we first consider the $N$ non-interacting fermions in a given, arbitrary, many-body
eigenstate of energy $E$ of ${\cal H}_N$ \eqref{HN}. We can conveniently label such a state by 
introducing the set of occupation numbers, denoted by $\{ n_k \}$, $k=0, 1, 2,\ldots$ with 
$n_k =0,1$, with $n_{k_1}=n_{k_2}= \ldots = n_{k_N}=1$ for the occupied single particle states and
$n_k=0$ otherwise. They satisfy the constraint $\sum_{k=0}^\infty n_k = N$. 
The corresponding many-body eigenfunction is given by a Slater determinant,
with the corresponding eigenenergy, 
\bea \label{manybodyeigen}
\displaystyle{\Psi_E({\bf x}) \equiv \Psi_{\{ n_k \}}({\bf x})= \frac{1}{\sqrt{N!}}  \det_{1\leq i,j \leq N} \phi_{k_i}(x_j)} \quad , \quad
E \equiv E_{\{ n_k \}} = \sum_{k=0}^\infty n_k \epsilon_k 
\eea 
with, e.g. $E=\mu_0 \,(k_1+k_2+\dots+k_N)$ for the harmonic oscillator. From now on
we use the shorthand notation ${\bf x} \equiv \{x_1,\cdots,x_N\}$. For convenience
we also use the quantum mechanical notation $\Psi_E({\bf x}) = \langle E|{\bf x} \rangle$. \\

To each eigenstate $|E \rangle$ we associate its $m$-time quantum JPDF
at time slices $\tau_1<\tau_2<\cdots< \tau_m$, with $\tau_i \in [0,\beta]$,
\bea
P_E({\bf x}^{(1)},\cdots, {\bf x}^{(m)}) = \frac{1}{B_{N,m}} \,
\langle E|{\bf x}^{(1)} \rangle \, \langle {\bf x}^{(1)} | e^{- (\tau_2-\tau_1) {\cal H}_N } | {\bf x}^{(2)} \rangle
\cdots \langle {\bf x}^{(m-1)} | e^{- (\tau_m-\tau_{m-1}) {\cal H}_N } | {\bf x}^{(m)} \rangle
\, \langle {\bf x}^{(m)} | E \rangle  \label{PEE} 
\eea 
where ${\cal H}_N$ is the $N$-body Hamiltonian defined in the previous sections. Here
\be
B_{N,m}= \langle E | e^{- (\tau_m-\tau_1) {\cal H}_N} | E \rangle = e^{- E (\tau_m-\tau_1)} \label{norm1} 
\ee 
is a normalization constant such that 
%{\red starting from here I have changed slice indices
%as $k \to \ell$ and $\ell \to \ell'$, please recheck carefully}
\be
\int  \left[ \prod_{i=1}^N \prod_{\ell=1}^m dx_i^{(\ell)} \right] \, P_E({\bf x}^{(1)},\cdots, {\bf x}^{(m)})  = 1
\ee
obtained by using the completeness relation $\int \prod_{i=1}^N dx_i^{(\ell)}  
 | {\bf x}^{(\ell)} \rangle \langle {\bf x}^{(\ell)} | = I$ where the $I$ is the identity in the $N$ body
Hilbert space. Note that $P_E$ is a symmetric function upon any permutation
of the particles at each given time, i.e. in each set $(x_1^{(\ell)},\cdots, x_N^{(\ell)})$ for each $\ell$.
Note also that for the case $m=1$ one recovers the usual one time quantum JPDF
\be
P_E({\bf x}) = |\langle E|{\bf x} \rangle|^2 = |\Psi_E({\bf x})|^2 \label{label} \;.
\ee

Let us note that the matrix elements in \eqref{PEE} are the $N$ fermion quantum propagator and can be written
as determinants of the quantum
propagator of the single particle problem
\bea
\langle {\bf x} | e^{- \tau {\cal H}_N } | {\bf y} \rangle = G^{(N)}(x_1,\cdots,x_N;\tau| y_1,\cdots,y_N;0)
&=& \frac{1}{N!} \det_{1\leq i,j \leq N} G(x_i,\tau|y_j,0) \label{detG2} 
\eea
as demonstrated in \eqref{detG}.

%
% given in \eqref{propdec},
%\bea
% \langle {\bf x} | e^{- \tau {\cal H}_N } | {\bf y} \rangle = \sum_E e^{- \tau E} \langle {\bf x} | E \rangle 
% \langle E | {\bf y} \rangle &=& \frac{1}{N!^2}  \sum_{k_1,k_2,\cdots ,k_N=0}^{+\infty} e^{- \sum_{\ell=1}^N \epsilon_{k_\ell} } 
% \det \phi_{k_i}(x_j)  \det \phi_{k_i}(y_j) \\
% &=& \frac{1}{N!} \det G(x_i,\tau|y_j,0) \label{detG} 
%\eea 
%where $G(x_i,\tau|y_j,0)$ is given in equation \eqref{propdec}. In obtaining the last
%equality, we have used the Cauchy-Binet formula for a discrete integration measure.

\subsection{Determinantal structure of the multi-time JPDF in a given fermionic eigenstate} 
\label{sec:det} 

We now define the correlation functions associated to the quantum probability measure \eqref{PEE}
for a fixed eigenstate of the energy $|E \rangle$. We first
recall the definitions and properties of the correlation functions at a fixed time. For this we set 
$m=1$ in \eqref{PEE}. Using $B_{N,1}=1$ in \eqref{norm1}, we obtain the JPDF as a determinant \cite{DLMS16}
\bea
P_E({\bf x}) = |\psi_E(x_1,\cdots,x_N)|^2 = \frac{1}{N!} | \det_{1\leq i,j \leq N} \phi_{k_i}(x_j)|^2 =  \frac{1}{N!}  
\det_{1\leq i,j \leq N} K(x_i,x_j;\{ n_k \}) \label{PE2} 
\eea 
in terms of the fixed eigenstate kernel 
\bea
K(x,y;\{ n_k \}) = \sum_{k=0}^{+\infty} n_k \phi^*_k(x) \phi_k(y) \;,
\eea 
where $n_k$ are the occupation numbers ($=0,1$) associated to the eigenstate $|E \rangle$, defined above. Note that this kernel has a self-reproducing property
\be
\int dz K(x,z;\{ n_k \})  K(z,y;\{ n_k \})  = K(x,y;\{ n_k \})  \label{srep} \;.
\ee
One now defines the $n$-point correlation functions $R_n$ in the eigenstate $|E \rangle$ as
\be
R_n(x_1,\cdots,x_n;\{ n_k \})  = \frac{N!}{(N-n)!} \int dx_{n+1} \cdots dx_N P_E({\bf x}) \;.
\ee
Using the determinantal form \eqref{PE2} for $P_E$ and the self-reproducing 
property \eqref{srep} one can show that all the $R_n$'s can be written as $n \times n$ determinants
\be
R_n(x_1,\cdots,x_n;\{ n_k \})  = \det_{1 \leq i,j \leq n} K(x_i,x_j;\{ n_k \}) \;.
\ee
This is usually referred to as the determinantal structure associated to non-interacting fermions. Accordingly, their
positions form a determinantal point process. In particular the density is given by
\be
N \rho_N(y) = R_1(y) = \Big\langle \sum_{i=1}^N \delta(x_i-y) \Big \rangle_E = K(y,y ; \{ n_k \}) = \sum_{k=0}^{+\infty} n_k
|\phi_k(y)|^2 \label{rho1} 
\ee
where $\langle \cdots \rangle_E$ denotes the average in the quantum state $|E \rangle$ i.e., with respect to (w.r.t.) \eqref{PE2}. \\

We now consider a number of time slices $m > 1$, in equation \eqref{PEE}. 
We recall that for $m=1$ we could write $P_E$ as a determinant (\ref{PE2}). The {\it theorem of
Eynard-Mehta} states that for any $m \geq 1$, $P_E$ can also be written as a 
$m N \times m N$ determinant as follows \cite{EM98,BorodinOlshanski}
\bea
P_E({\bf x}^{(1)},\cdots, {\bf x}^{(m)}) = \frac{1}{N!^m} \det_{1 \leq i,j \leq N, 1 \leq \ell,\ell' \leq m} 
K( x^{(\ell)}_i , \tau_{\ell} ; x^{(\ell')}_j , \tau_{\ell'} ;\{ n_k \})  \label{EM} 
\eea 
in terms of the {\it extended kernel} given by 
%{\red I have not changed indices unless strictly needed so
%I advise to keep $\tau_i$ and $\tau_j$ in all kernel formula below, even if slightly misleading... 
%too much risk of error for little gain..what do you think?}
\begin{eqnarray}
&&K(x,\tau_i; y,\tau_j;\{ n_k \}) = \sum_{k=0}^\infty n_k \, e^{-\epsilon_k(\tau_j - \tau_i)} \phi^*_k(x) \phi_k(y) \; \; \;, \; \tau_i \geq \tau_j 
\label{Kext10}
\\
&&K(x,\tau_i; y, \tau_j;\{ n_k \}) = - \sum_{k=0}^\infty (1-n_k) e^{-\epsilon_k (\tau_j-\tau_i) }\phi^*_k(x) \phi_k(y) \; \; \;, \; \tau_i < \tau_j \;.
\label{Kext20} 
\end{eqnarray}
Note that in the original EM theorem the kernel is not expressed explicitly 
in terms of the occupation numbers $n_k$. Here we re-expressed this kernel in terms of
the $n_k$ variables using the formulation of Ref. \cite{BorodinOlshanski} as detailed
in the Appendix \ref{app:details}. We found this formulation of the extended
kernel the most convenient way to generalize to finite temperature (see next section).
Note that for the case $m=1$ we only need to consider the first equation (\ref{Kext10}) since $\tau_i=\tau_j$, and
we recover the known result for $m=1$ in \eqref{PE2}. Note that the ordering in \eqref{Kext10}-\eqref{Kext20} concerns only the slice index $j$.
Although we used it in a situation where the variables $\tau_i$ are in increasing order,
the determinantal formula extends formally to a more general case (as used below in the study of real time dynamics). 

The Eynard-Metha theorem in fact guarantees that there is {\it an extended determinantal structure} for all
spatio-temporal correlations: upon integration of (\ref{EM}) over any subset of the space 
coordinates $x^{(\ell)}_i$ at various times, the corresponding marginal probability remains a determinant with the
same kernel. In particular consider the natural generalization of the density involving several times, i.e. 
the following correlation function
\bea
\hat R_m(y_1,\tau_1 ; \cdots,y_m,\tau_m;\{ n_k \})  = \Big \langle \sum_{i=1}^N \delta(x^{(1)}_i - y_1) \cdots \sum_{j=1}^N \delta(x^{(m)}_j- y_m) \Big \rangle_E  \label{Rm1}
\eea 
which is by definition symmetric under the exchange of any pair $(y_i,\tau_i)$ and $(y_j,\tau_j)$.
For a given ordering of the times $\tau_1 < \cdots < \tau_m$,
it can be expressed as an $m \times m$ determinant using the EM theorem
\bea
\hat R_m(y_1,\tau_1 ; \cdots,y_m,\tau_m;\{ n_k \})  = \det_{1 \leq \ell,\ell' \leq m} 
K(y_{\ell} , \tau_{\ell} ; y_{\ell'} , \tau_{\ell'} ;\{ n_k \}) \;, \label{Rm2}
\eea 
where $K$ is given in \eqref{Kext10}-\eqref{Kext20}.
For instance, for $m=2$, the two time density-density correlation reads, for $\tau_1<\tau_2$
\be
\hat R_2(y_1;\tau_1,y_2;\tau_2;\{ n_k \})  = \sum_p n_p |\phi_p(y_1)|^2  \sum_k n_k |\phi_k(y_2)|^2 
+ \sum_{p,k} (1-n_p) n_k \phi^*_p(y_1) \phi_p(y_2) \phi_k(y_1) \phi^*_k(y_2)
e^{- (\epsilon_p -\epsilon_k) (\tau_2-\tau_1)} \label{2time} \;.
\ee
By integration over $y_1$ one finds $N R_1(y_2)$. For $\tau_2=\tau_1$ one finds instead
\bea
\hat R_2(y_1;\tau_1,y_2;\tau_1;\{ n_k \})  &=& \sum_p n_p |\phi_p(y_1)|^2  \sum_k n_k |\phi_k(y_2)|^2 
- \sum_{p,k} n_p n_k \phi_p(y_1) \phi_p^*(y_2) \phi^*_k(y_1) \phi_k(y_2) 
\\
&=& \det_{1 \leq i,j \leq 2} K(y_i,y_j;\{ n_k \})  = R_2(y_1,y_2) \;, 
\eea
recovering the formula for the two point same time correlation function $R_2$. 
Similar determinantal formulas are available for more general space time correlation functions,
as detailed in the Appendix \ref{app:details}, where the self-reproducing properties of the extended kernel are also discussed.  

One consequence of the determinantal structure is that one can express as Fredholm determinants
averages of the form \cite{Borodin1,Johansson2005}
\bea
\langle \prod_{\ell=1}^m \prod_{i=1}^N \left( 1 + g_{\ell}(x_i^{(\ell)}) \right) \rangle_E = {\rm Det} [ I + g {\cal K} ] 
\eea 
where $(g {\cal K})_{\ell,\ell'}(x,y)$ with $1 \leq \ell,\ell' \leq m$ is the matrix kernel 
\be
(g {\cal K})_{\ell,\ell'}(x,y) = g_{\ell}(x) K(x, \tau_{\ell} ; y , \tau_{\ell'} ;\{ n_k \}) \;.
\ee 

\begin{figure}[ht]
\includegraphics[width=0.6\linewidth]{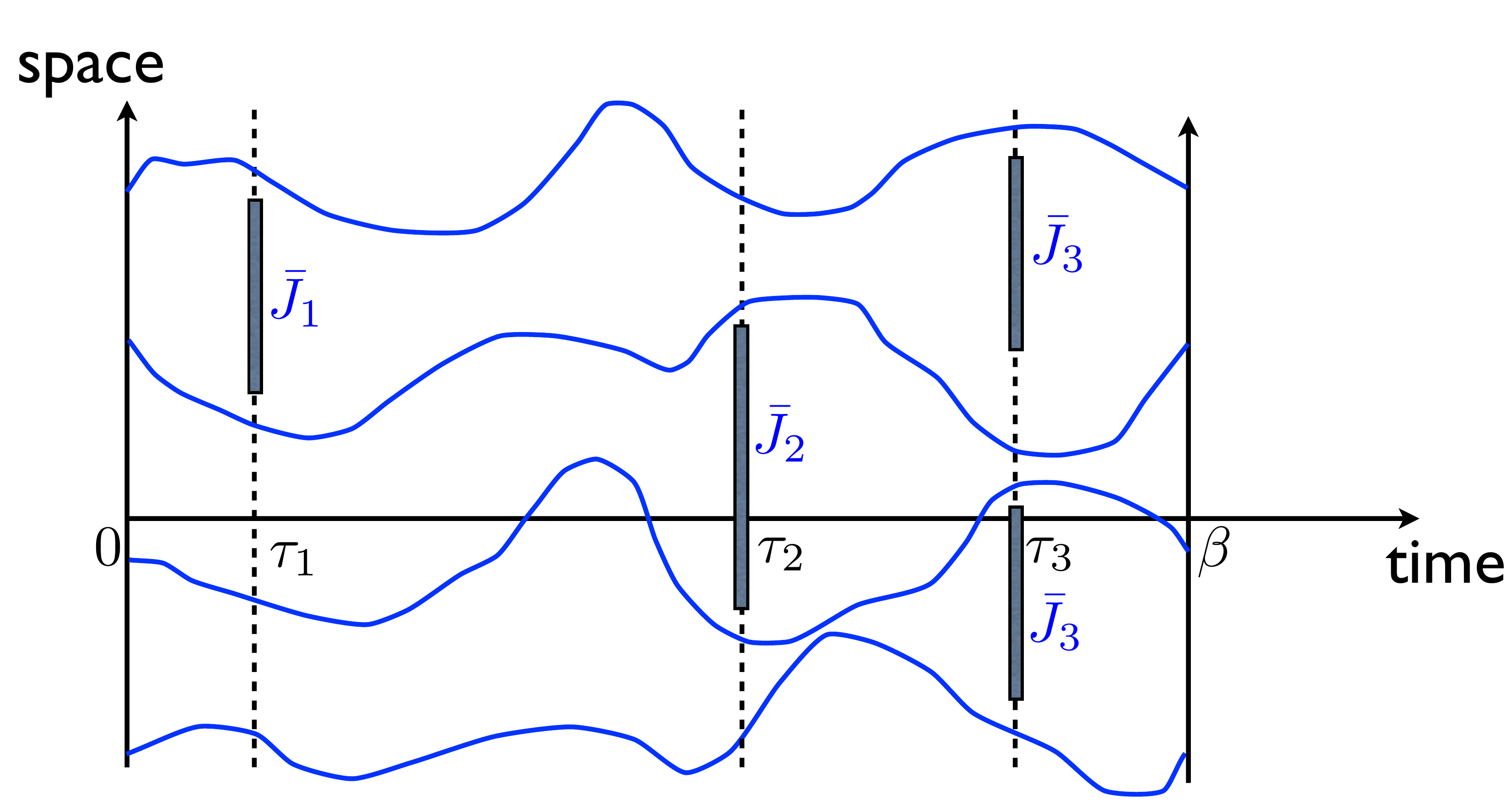}
\caption{Sketch of the hole probability for $N=4$ fermions and $m=3$ different times. Note the periodic boundary condition in the time direction.}\label{Fig_hole_proba}
\end{figure}

Let us choose now $g_{\ell}(x) = P_{J_{\ell}}(x) - 1 = - P_{\bar J_{\ell}}(x)$ where $P_J(x)$ denotes the 
indicator function (i.e., projector) on the subset $J$ of $\mathbb{R}$, with $P_j(x)=1$ if $x \in J$ and 
$P_j(x)=0$ otherwise, and $\bar J$ denotes the complementary subset. 
This allows to express the generalized multi-time "hole probabilities" (see Fig. \ref{Fig_hole_proba})  as
\bea
&& {\rm Prob}( x_i^{(\ell)} \in J_{\ell} ; i =1, \cdots, N ; \ell=1,\cdots, m) = 
{\rm Det} [ I - P {\cal K} ]  \\
&& (P {\cal K})_{\ell,\ell'}(x,y) = P_{J_{\ell}}(x) K(x, \tau_{\ell} ; y , \tau_{\ell'} ;\{ n_k \}) \;,
\eea 
which generalizes the standard hole probability for $m=1$.
%{\red show 2 figures, one with the $J_k$ and one with the $\bar J_k$.}

\section{Application of the EM theorem: extended kernels at the bulk and the edge of the ground state of $N$ fermions in a harmonic trap}
\label{sec:ground} 

In this section we specialize to the case where $|E \rangle$ is the
ground state, of wave-function denoted $|\Psi_0\rangle$. We apply the
results the previous calculation to study the imaginary time dynamics at the bulk and the edge of the
Fermi gas at $T=0$. The occupation numbers are
$n_k=1$ for $k=0,\cdots,N-1$ and $n_k=0$ for $k \geq N$. This leads to
the JPDF of the positions of the fermions at equal time as
\bea
P_0({\bf x}) = |\psi_0(x_1,\cdots,x_N)|^2 =  \frac{1}{N!}  
\det_{1\leq i,j \leq N} K_N(x_i,x_j) \label{PE4} \;,
\eea 
in terms of the GUE kernel 
\be
K_N(x,y) = \sum_{k=0}^{N-1} \phi^*_k(x) \phi_k(y) \;.
\ee 
Note that Eq. (\ref{PE4}) identifies with the GUE measure in (\ref{GibbsN}), 
see Ref. \cite{DLMS16} (section III). 

Let us now discuss the large $N$ limit. In that limit there are
asymptotic Plancherel-Rotach formula (see for instance Ref. \cite{For10}) for the eigenfunctions
$\phi_k(x)$ of the harmonic oscillator, which allow to
describe the system at large $N$. As is well known 
the density $\rho_N(x)$, Eq. (\ref{rho1}), converges to the Wigner semi-circle law
\bea
\rho_N(x) = \frac{\alpha}{\sqrt{N}} f_W\left(\frac{\alpha x}{\sqrt{N}}\right) \quad , \quad f_W(z) = \frac{1}{\pi} \sqrt{2-z^2} \;. 
\label{rhonx}
\eea 
It exhibits a sharp edge $x_{\rm edge}= \sqrt{2 N}/\alpha$, where here $\alpha=\sqrt{\omega}$ as
we chose $\hbar=m=1$, such that for $N \to +\infty$ the density vanishes
outside the interval $[-x_{\rm edge},x_{\rm edge}]$. It is well known (see \cite{DLMS16} for a recent review)
that there are thus two regions of
interest:
\begin{itemize}

\item{
In the bulk, the density is finite and one can define a typical interparticle spacing at point $x$
\bea
\ell_N(x) = \frac{2}{\pi N \rho_N(x)} \label{lnx} \;.
\eea 
On that scale, the kernel exhibits the following scaling form in the
large $N$ limit, for $|x-y|= {\cal O}(\ell_N(x))$, in terms of the sine-kernel
\bea
K_N(x,y) \simeq \frac{1}{\ell_N(x)} {\cal K}^{\rm bulk}\left(\frac{|x-y|}{\ell_N(x)}\right)  \quad , \quad 
{\cal K}^{\rm bulk}(z) = \frac{\sin( 2 z)}{\pi z} \;.
\eea 
}

\item{
Near the edge $x_{\rm edge}$, the density vanishes and at finite $N$ there are strong quantum 
fluctuations.
The density is smeared on a scale of order 
\bea
w_N = \frac{1}{\alpha \sqrt{2}} N^{-1/6} \label{wN}
\eea 
and the kernel, in the region $|x-x_{\rm edge}| = {\cal O}(w_N)$, $|y-x_{\rm edge}| = {\cal O}(w_N)$, 
takes the scaling form
\bea
K_N(x,y) \simeq \frac{1}{w_N} {\cal K}^{\rm edge}\left(\frac{x-x_{\rm edge}}{w_N}, \frac{y-x_{\rm edge}}{w_N}\right)
\eea 
in terms of the Airy kernel
\bea
{\cal K}^{\rm edge}(s,s') = \int_0^{+\infty} dv {\rm Ai}(s+v) {\rm Ai}(s'+v) \;.
\eea 
}

\end{itemize}

Let us now discuss the multi-time generalization of these results. Specializing the formulas 
(\ref{EM})-(\ref{Kext20}) to the ground state, we obtain that the
multi-time correlations are given as determinants in terms of the
so called {\it extended Hermite kernel} 
\bea
&&K_N(x,\tau_i; y,\tau_j) = \sum_{k=0}^{N-1} e^{- k \omega (\tau_j - \tau_i)} \phi^*_k(x) \phi_k(y) \; \; \;, \; \tau_i \geq \tau_j 
\label{KHext1}
\\
&&K_N(x,\tau_i; y,\tau_j) = - \sum_{k=N}^\infty e^{- k \omega (\tau_j-\tau_i) }\phi^*_k(x) \phi_k(y) \; \; \;, \; \tau_i < \tau_j 
\label{KHext2} \;.
\eea
In the large $N$ limit the asymptotics of this kernel has been obtained, and exhibit again two regimes:

\begin{itemize}

\item{
In the bulk, one finds the {\it extended sine-kernel} (see for instance Ref. \cite{AM04}, Eqs. (7.1) and (7.11))
\bea
&& K_N(x,\tau_i; y,\tau_j) = \frac{1}{\ell_N(x)} {\cal K}^{\rm bulk}\left(\frac{|x-y|}{\ell_N(x)}, \frac{\tau_i-\tau_j}{\ell_N(x)^2}\right) 
\label{extendedsine0} 
\\
&& {\cal K}^{\rm bulk}(z,\tau) = \frac{1}{\pi} \int_0^{2} e^{v^2 \tau/2} \cos(v z) dv \quad , \quad \tau \geq 0 \\
&&  {\cal K}^{\rm bulk}(z,\tau) = - \frac{1}{\pi} \int_2^{+\infty} e^{v^2 \tau/2} \cos(v z) dv \quad , \quad \tau < 0 \;.
\eea
Note that there is a global factor $e^{ \frac{1}{2} \omega^2 x^2 (\tau_i-\tau_j)}$ in \eqref{extendedsine0} 
which we discarded since it drops in any correlation function. }

\item{
Near the edge $x_{\rm edge}=\sqrt{2 N}$, in the region $|x-x_{\rm edge}| = {\cal O}(w_N)$, $|y-x_{\rm edge}| = {\cal O}(w_N)$ 
one finds the {\it extended Airy kernel} with the scaling form (see Ref.~\cite{Mac94} and Ref.~\cite{AM04} Eq. (7.1) and (7.3))
\bea
K_N(x,\tau_i;y, \tau_j)  \simeq \frac{1}{w_N} {\cal K}^{\rm edge}(\frac{x-x_{\rm edge}}{w_N}, 
\frac{y-x_{\rm edge}}{w_N}, (\tau_i - \tau_j) \omega N^{1/3}) \label{extendededge0}
\eea 
in terms of the Airy kernel
\bea
&& {\cal K}^{\rm edge}(s,s',u) = \int_0^{+\infty} dv e^{- v u} \Ai(s+v) \Ai(s'+v) \quad , \quad u \geq 0 \\
&& {\cal K}^{\rm edge}(s,s',u) = - \int_{-\infty}^0 dv e^{- v u} \Ai(s+v) \Ai(s'+v) \quad , \quad u < 0 \;.
\eea 
The third (time) argument of the scaling function ${\cal K}^{\rm edge}$ reflects that in the scaling region the time $\tau$ scales
as $N^{-1/3}$ for large $N$. This is a manifestation of the local Brownian scaling with $\tau \sim w_N^2$ where
$w_N \sim N^{-1/6}$ [see Eq. \eqref{wN}] is the relevant length scale at the edge.
Note that there is a global factor $e^{ N \omega  (\tau_i-\tau_j)}$ in \eqref{extendededge0}
which we discarded since it drops in any correlation function. 
This extended Airy kernel also describes the correlations of the Airy$_2$ process, as
discussed in the introduction, and will be discussed in more details below.

}

\end{itemize}

\section{Application of the EM Theorem: generalization to finite temperature}
\label{sec:finiteT} 

%In the preceding section we introduced $P_T(x_1,\cdots,x_N)$,
%which, up to a normalization constant in  \eqref{PTxN3}, is the one-time JPDF of $N$ particle time-periodic non-crossing
%OU processes, as well as the one imaginary time quantum JPDF of
%$N$ fermions in a harmonic well. 
%

\subsection{General framework}

We now consider $N$ fermions in a harmonic potential in the canonical ensemble at
finite temperature $T=1/\beta$, as described in section \ref{sec:Nfermions}. 
By analogy with the one-time finite temperature JPDF in \eqref{PTxN}, \eqref{PTxN2} and \eqref{label},
we define the canonical multi-time JPDF at finite temperature as
\bea
\tilde P_{\beta}({\bf x}^{(1)},\cdots, {\bf x}^{(m)}) = \frac{1}{Z_N(\beta)} \sum_{E} P_E({\bf x}^{(1)},\cdots, {\bf x}^{(m)}) e^{- \beta E} \;,
\eea
where $P_E({\bf x}^{(1)},\cdots, {\bf x}^{(m)})$ is given in \eqref{PEE}. Note that for $m=1$,
$\tilde P_{\beta}({\bf x})=P_T(x_1,\cdots,x_N)$ defined in \eqref{PTxN2}. Using \eqref{PEE} it is easy to rewrite this
finite temperature JPDF as
\be
\tilde P_{\beta}({\bf x}^{(1)},\cdots, {\bf x}^{(m)}) = \frac{1}{Z_N(\beta)} 
 \langle {\bf x}^{(1)} | e^{- (\tau_2-\tau_1) {\cal H}_N } | {\bf x}^{(2)} \rangle
\cdots \langle {\bf x}^{(m-1)} | e^{- (\tau_m-\tau_{m-1}) {\cal H}_N } | {\bf x}^{(m)} \rangle
\,  \langle {\bf x}^{(m)} | e^{- (\beta-(\tau_m-\tau_{1}) ) {\cal H}_N } | {\bf x}^{(1)} \rangle \label{Pbeta} \;.
\ee
Note that $\tilde P_{\beta}$ is a symmetric function under any permutation
in each set $(x_1^{(\ell)},\cdots, x_N^{(\ell)})$ for each $\ell$. It is normalized to
unity upon integration of the coordinates over $\mathbb{R}^{N m}$. 
Note that integrating over all time slices except one leads to 
\be
\int d{\bf x}^{(2)} \cdots d{\bf x}^{(m)}  \tilde P_{\beta}({\bf x}^{(1)},\cdots, {\bf x}^{(m)}) = \frac{1}{Z_N(\beta)} 
\langle {\bf x}^{(1)} | e^{- \beta {\cal H}_N } | {\bf x}^{(1)} \rangle = P_T(x_1^{(1)},\cdots, x_N^{(1)})
\ee
where $P_T$, given in \eqref{PTxN}, is the equilibrium JPDF at a single time. 

Although this canonical JPDF \eqref{Pbeta} for arbitrary $N$ is a product of determinants, see Eq. (\ref{detG}), 
and is thus itself a determinant, the associated point process is not determinantal as
correlation functions (i.e obtained by integrating over some of the variables) are not themselves
determinants. To preserve the determinal structure, as was noted in the case $m=1$ \cite{DLMS15,DLMS16,Joh07}
it is necessary to study the problem in the grand canonical ensemble where
the number of fermions $N$ fluctuates. One defines the grand canonical partition
function as
\be
{\cal Z}(\beta,\mu) = \sum_{E,N} e^{- \beta E - \mu N}  \;.
\ee
In the large $N$ limit the averages of physical observables, such as correlation functions,
become identical in the canonical and grand canonical ensembles.

Note that the total number of fermions and energy are, respectively, $N= \sum_{k \geq 0} n_k$ and 
$E= \sum_{k \geq 0} n_k \epsilon_k $
where $n_k=0,1$ depending on whether the $k$-th single particle level is empty or
occupied. Consequently, in the grand canonical ensemble the occupation numbers 
$n_k$ are i.i.d. Bernoulli random variables. We now define the multi-correlations at finite temperature both in the canonical and
grand canonical ensembles. Here we give only the multi-time density correlation. 
The canonical correlation is
\bea
\hat R_m(y_1,\tau_1 ; \cdots;y_m,\tau_m;\beta)  &=& \frac{1}{Z_N(\beta)} \sum_E e^{-\beta E} 
\Big \langle \sum_{i=1}^N \delta(x^{(1)}_i - y_1) \cdots \sum_{j=1}^N \delta(x^{(m)}_j- y_m) \Big \rangle_E \\
&=&  \frac{1}{Z_N(\beta)} \sum_{\{n_k\}} \left[
\det_{1 \leq \ell,\ell' \leq m} K(y_{\ell} , \tau_{\ell} ; y_{\ell'} , \tau_{\ell'} ;\{ n_k \}) e^{- \beta \sum_{k \geq 0} n_k \epsilon_k} 
\delta\left(\sum_{k \geq 0} n_k,N\right) \right]
\eea 
where the last delta is a Kronecker delta function, while the grand canonical correlation is
\bea
\hat R^G_m(y_1,\tau_1 ; \cdots;y_m,\tau_m;\beta,\mu)  &=& \frac{1}{{\cal Z}(\beta,\mu)} \sum_{E,N} e^{-\beta E-\mu N } 
\Big \langle \sum_{i=1}^N \delta(x^{(1)}_i - y_1) \cdots \sum_{j=1}^N \delta(x^{(m)}_j- y_m) \Big \rangle_E \\
&=& \frac{1}{{\cal Z}(\beta,\mu)} \sum_{\{n_k\}} \left[
\det_{1 \leq \ell,\ell' \leq m} K(y_{\ell} , \tau_{\ell} ; y_{\ell'} , \tau_{\ell'} ;\{ n_k \})
%\det_{1 \leq k,\ell \leq m} K(y_k , \tau_k ; y_\ell , \tau_\ell ;\{ n_k \})) 
e^{- \beta \sum_{k \geq 0} n_k \epsilon_k - \mu \sum_{k \geq 0} n_k} \right]
\eea 
where, in these formula, the superscript `G' refers to the grand-canonical ensemble and $K$ is the extended kernel given in \eqref{Kext10}-\eqref{Kext20}. 

We now use the property that
\bea
\Big \langle \det_{1 \leq \ell ,\ell' \leq m} K(y_{\ell} , \tau_{\ell} ; y_{\ell'} , \tau_{\ell'} ;\{ n_k \}) \Big \rangle
= \det_{1 \leq \ell,\ell' \leq m} K(y_{\ell} , \tau_{\ell} ; y_{\ell'} , \tau_{\ell'} ;\{ \langle n_k \rangle \}) \;,
\eea 
which holds for any averaging such that the variable 
$n_k$ are independent. Note that we also use the linearity of $K$ in the $n_k$
(see \cite{DLMS16} for a proof). This property can be used in the
grand canonical ensemble with 
\be
\langle n_k \rangle = \frac{1}{e^{\beta (\epsilon_k - \mu)} +1 } \;. \label{fermi_factor}
\ee
Hence the grand canonical correlation can be written as the following
determinant 
\bea
\hat R^G_m(y_1,\tau_1 ; \cdots;y_m,\tau_m;\beta,\mu)  = \det_{1 \leq \ell,\ell' \leq m} 
K(y_{\ell} , \tau_{\ell} ; y_{\ell'} , \tau_{\ell'} ;\beta,\mu) \label{RG} 
\eea 
where the extended, grand canonical kernel, obtained by replacing $n_k$ by
$\langle n_k \rangle$ in \eqref{Kext10}-\eqref{Kext20} is
\begin{eqnarray}
&&K(x,\tau_i; y,\tau_j ;\beta,\mu) = \begin{cases} & \sum_{k=0}^\infty \frac{e^{(\epsilon_k-\mu)(\tau_i - \tau_j)} }{e^{\beta (\epsilon_k - \mu)} +1 }  \,  \phi^*_k(x) \phi_k(y) \; \; \;, \; \tau_i \geq \tau_j 
\label{Kext1}
\\
& \\
&
%K(\tau_i,x; \tau_j,y; \beta,\mu) = 
- \sum_{k=0}^\infty 
\frac{e^{(\epsilon_k-\mu)(\tau_i - \tau_j)} }{e^{-\beta (\epsilon_k - \mu)} +1 }  
\phi^*_k(x) \phi_k(y) \; \; \;, \, \tau_i < \tau_j \end{cases} 
\label{Kext2} 
\end{eqnarray}
where for convenience we have inserted in the kernel a global factor $e^{- \mu (\tau_i - \tau_j)}$ which cancels
out in any correlation function. The extended kernel is a function only of the time difference
$\tau_i-\tau_j$ and is not continuous at $\tau_i=\tau_j$ where it has a jump
\bea
[ K(x,\tau_i; y,\tau_j ;\beta,\mu) ]_{\tau_i=\tau_j^+} - [ K(x,\tau_i; y,\tau_j;\beta,\mu) ]_{\tau_i=\tau_j^-} 
= \delta(x-y) \;.  \label{jump} 
\eea 
Note that the second form is obtained from the first one
by replacing $\tau_i - \tau_j \to \beta + \tau_i - \tau_j$, more precisely
\bea\label{anti}
K(x,\tau_i;y, \tau_j ;\beta,\mu) = - K(x,\tau_i+\beta; y, \tau_j ;\beta,\mu) \quad , \quad \tau_i < \tau_j \quad \mbox{and} \quad 
\tau_i + \beta \geq \tau_j  \;.
\eea 
Hence the extended kernel is an anti-periodic in each time variable with
period $\beta$. As a consequence it is sufficient to consider the fundamental domain
$(\tau_i, \tau_j) \in [0,\beta]\times [0,\beta]$. For any other value of $(\tau_i, \tau_j)$ one can
construct the kernel using the anti-periodicity property (\ref{anti}). 

In the fermionic literature this extended kernel is known as the ``temperature
Green's function'' which however is often studied for non-interacting fermions
in absence of external confining potential (i.e., using plane waves as single particle
eigenfunctions), see for instance Ref. \cite{FW71} p. 233 Chapter 7 formula (23.30)
where we can identify $K(x,\tau_i; y,\tau_j ;\beta,\mu)$
with ${\cal G}^0(y,\tau_j;x,\tau_i)$ given there in the absence of the confining
potential - note the order of the arguments.
In that context, determinantal formulas
such as (\ref{RG}) can be seen as consequences of the Wick theorem~\cite{gaudin} extended
to non-equal times and arbitrary confining potentials, 
although they are not often explicitly stated as such in the condensed matter literature. 

\subsection{Bulk and edge regime in the harmonic trap at finite temperature}
\label{sec:bulkedge}

We now derive the expressions for the extended kernel in the limit of large $N$ 
both in the bulk and at the edge. We do not give details since the derivation is
a simple combination of the results of \cite{DLMS15,DLMS16} and the zero temperature results discussed
above. 

\begin{itemize}

\item{
In the bulk, the characteristic temperature scale is $N \omega$, which is the Fermi energy, hence one introduces,
as in \cite{DLMS15,DLMS16}, the scaling variable 
\be
{\sf y} = \beta N \omega \label{sfy} \;.
\ee 
The relation (\ref{fermi_factor}) implies that the chemical potential is related to the mean number of
particles $N$ as $e^{\beta \mu}=e^{{\sf y}}-1$. 
Then one finds a finite temperature generalization of the extended sine-kernel 
\bea
&& K(x,\tau_i; y,\tau_j;\beta, \mu) = \frac{1}{\ell_N(x)} {\cal K}_{\sf y}^{\rm bulk}\left(\frac{|x-y|}{\ell_N(x)}, \frac{\tau_i-\tau_j}{\ell_N(x)^2}\right) 
\label{extendedsine} 
\\
&& {\cal K}_{\sf y}^{\rm bulk}(z,\tau) = \begin{cases}& \frac{1}{\pi} \int_0^{+\infty} \label{extendedsine2}
e^{v^2 \tau/2} \frac{\cos(v z)}{ 1+ \frac{e^{{\sf y} v^2/4}}{e^{\sf y}-1}} dv \quad , \quad 0 \leq \tau < \tau_\beta \\
& \\
%&&  {\cal K}_{\sf y}^{\rm bulk}(z,\tau) = 
& - \frac{1}{\pi} \int_0^{+\infty} 
e^{v^2 \tau/2} \frac{\cos(v z)}{1+ e^{- {\sf y} v^2/4} (e^{\sf y}-1)} dv \quad , \quad - \tau_\beta \leq \tau < 0 
\end{cases} \; \label{extendedsine3}  \\
&& \tau_\beta=\beta/\ell_N(x)^2 = \frac{{\sf y}}{2} (1- \omega x^2/(2N)) \;.
\label{taubeta}
\eea
 }
To derive \eqref{taubeta} we have used equation 
\eqref{lnx} for $\ell_N(x)$ and the expression for the density $\rho_N(x)$ in equation \eqref{rhonx}.
Note that as we discussed in the previous section the kernel $K(\tau_i,x; \tau_j,y;\beta, \mu)$
is antiperiodic with period $\beta$~(\ref{anti}). Hence the scaled kernel in Eqs. \eqref{extendedsine2}, 
\eqref{extendedsine3} is antiperiodic in the variable $\tau$ with period $\tau_\beta$
%=\beta/\ell_N(x) = \frac{y}{2} (1- \omega x^2/(2N))$ 
given by \eqref{taubeta} in the chosen rescaled units. 
It is easy to check that 
\be
{\cal K}^{\rm bulk}_y(z,\tau=0^+)- {\cal K}^{\rm bulk}_y(z,\tau=0^-) = \delta(z) \;,
\ee 
in agreement with the general property \eqref{jump}.
Since $\tau_\beta \leq y/2$ one can check that the integrals in \eqref{extendedsine3} are always convergent, and that the
process is thus well defined. 

\item{
Near the edge $x_{\rm edge}=\sqrt{2 N/\omega}$, in the region $|x-x_{\rm edge}| = {\cal O}(w_N)$, $|y-x_{\rm edge}| = {\cal O}(w_N)$ 
the relevant temperature scale is $N^{1/3} \omega$, hence one defines
\be
b = \beta \omega N^{1/3} \;.
\ee 
Then one finds the finite temperature extended Airy kernel, with the scaling form 
\bea
K(x,\tau_i; y,\tau_j;\beta,\mu)  \simeq \frac{1}{w_N} {\cal K}^{\rm edge}_b\left(\frac{x-x_{\rm edge}}{w_N}, 
\frac{y-x_{\rm edge}}{w_N}, (\tau_i - \tau_j) \omega N^{1/3}\right) \label{scaledge} 
\eea 
in terms of the finite temperature extended Airy kernel
\bea
&& {\cal K}_b^{\rm edge}(s,s',u) = \begin{cases} & \int_{-\infty}^{+\infty} dv \frac{e^{- u v}}{ e^{- b v} +1 } \Ai(s+v) \Ai(s'+v) \quad , \quad 0 \leq u < b \label{TextAiry0}  \\
& \\
&
% {\cal K}_b^{\rm edge}(s,s',u) =
 - \int_{-\infty}^{+\infty} dv \frac{e^{- u v}}{e^{b v} +1} \Ai(s+v) \Ai(s'+v) \quad , \quad - b \leq u < 0 \end{cases}  \label{TextAiry1} \;.
\eea 
The same remark as above holds for the antiperiodicity of the kernel. The rescaled
kernel is now antiperiodic in the variable $u$ with period $b$. Again one easily checks that 
\be
{\cal K}^{\rm edge}_y(s,s',u=0^+)- {\cal K}^{\rm edge}_y(s,s',u=0^-) = \delta(s-s') 
\ee 
in agreement with the general property \eqref{jump}.

}

\end{itemize}

\section{Relation to multi-time correlation functions in the OU and finite temperature Airy process}
\label{sec:TAiry} 

\subsection{Link between fermions at finite temperature and non-intersecting time-periodic OU processes} 

\begin{figure}
\includegraphics[width = 0.6\linewidth]{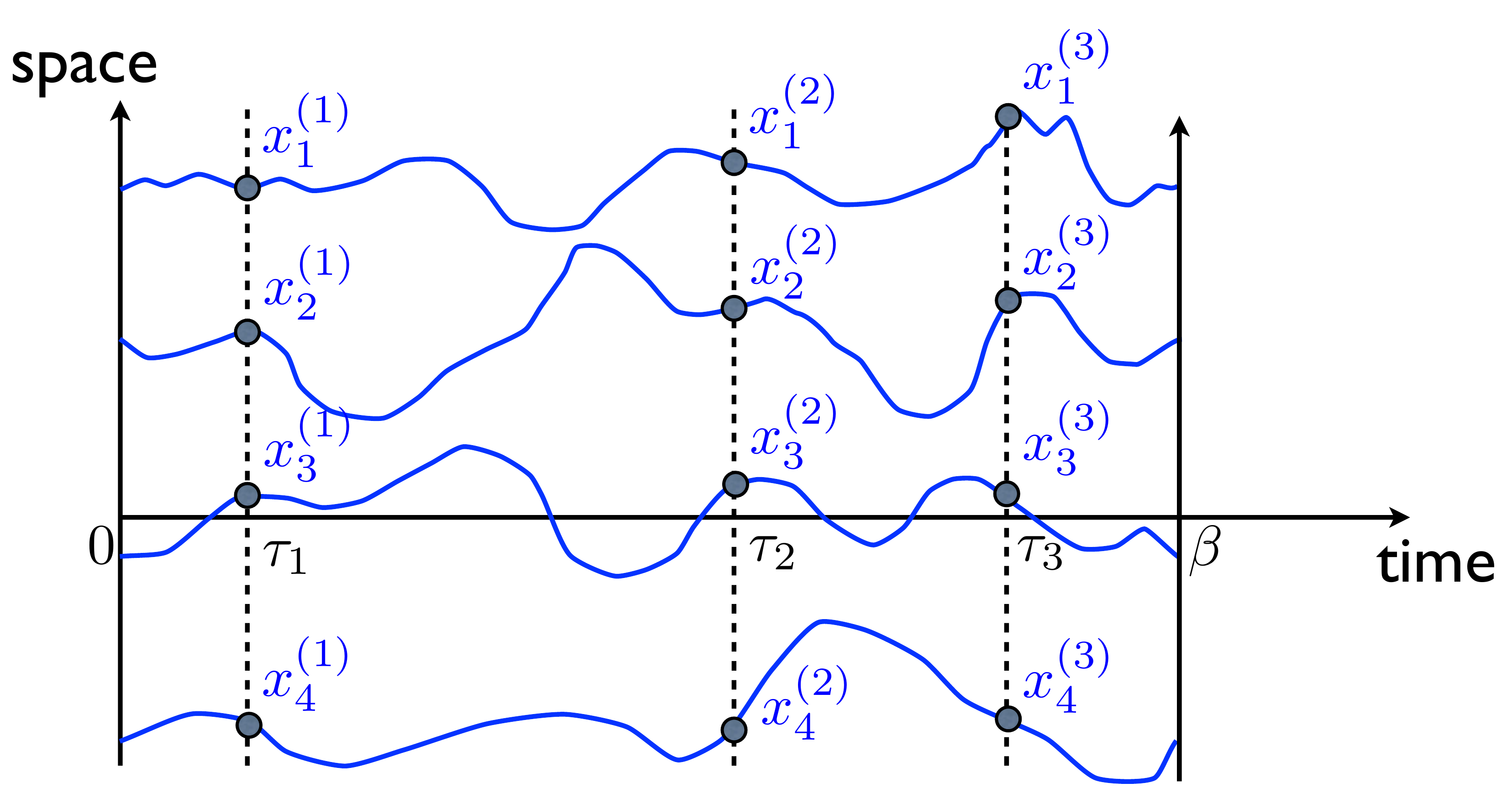}
\caption{Sketch of a trajectory of $N=4$ non-intersecting time-periodic OU processes contributing to the multi-time correlation function $\tilde P_{\beta}({\bf x}^{(1)}, {\bf x}^{(2)}, {\bf x}^{(3)})$, i.e., with $m=3$ [see Eq. (\ref{94})].}\label{fig_multi_time}
\end{figure}

Here we discuss the connection between the multi-time JPDF of fermions at finite temperature and the time-periodic $N$ non-intersecting OU processes. The one-time case was studied before in 
section \ref{sec:Nfermions}. We start from the multi-time JPDF of the fermions
$\tilde P_{\beta}({\bf x}^{(1)},\cdots, {\bf x}^{(m)})$ defined in Eq. (\ref{Pbeta}). Each term in the
$m$-fold product can be interpreted in terms of the OU propagator.
%Let us define as $P^{(N)}_{\rm OU}(x_1,\cdots,x_N;\tau| y_1,\cdots,y_N,0)$
%the probability that $N$ OU processes, starting initially from ordered
%positions ${\bf y}= y_1> \cdots > y_N$ at time zero, arrive at final positions with the same order
%${\bf x}= x_1>  \cdots > x_N$ at time
%$\tau$, and remain non-intersecting in the time interval $[0,\tau]$. 
%From the Karlin McGregor formula mentioned earlier in section \ref{sec:Nfermions}, we can write
%\bea
% P^{(N)}_{\rm OU}(x_1,\cdots,x_N;\tau| y_1,\cdots,y_N,0)
%& = & \det P_{\rm OU}(x_i , \tau | y_j, 0) \\
%&= & \det [ e^{- \mu_0 \frac{x_i^2}{2}} G(x_i , \tau | y_j, 0) e^{\mu_0 \frac{y_i^2}{2}} ] \\
%& = & N! \,  e^{- \mu_0 \sum_{i=1}^N \frac{x_i^2}{2}}
%\langle {\bf x} | e^{- \tau {\cal H}_N } | {\bf y} \rangle e^{ \mu_0 \sum_{i=1}^N \frac{y_i^2}{2}}
%\eea
%where in the second line we have used the correspondence \eqref{correspondence1}
%between the single-particle OU propagator and the quantum propagator of the fermions in a harmonic potential in imaginary time. In the last line we have used Eq. (\ref{detG}). 
Using Eqs. \eqref{GN} and \eqref{detG} we can thus rewrite Eq. (\ref{Pbeta}) as
\be
\tilde P_{\beta}({\bf x}^{(1)},\cdots, {\bf x}^{(m)}) = \frac{1}{(N!)^m Z_N(\beta)}
\left[ \prod_{\ell=1}^{m-1} 
P^{(N)}_{\rm OU}({\bf x}^{(\ell+1)},\tau_{\ell+1}|{\bf x}^{(\ell)}, \tau_\ell) \right] \times 
P^{(N)}_{\rm OU}({\bf x}^{(1)}, \tau_1+\beta|{\bf x}^{(m)},\tau_m) \label{94} \;,
\ee
where in this case, at variance with Eq. (\ref{Pbeta}), each ${\bf x}^{(\ell)}= x^{(\ell)}_1>  \cdots > x^{(\ell)}_N$
is an ordered set of coordinates, and the times are ordered as $0 \leq \tau_1 < \tau_2 < \cdots < \tau_N \leq \beta$. We can therefore see the right hand side as the probability that the time-periodic non-intersecting
OU process, wrapped on a cylinder of perimeter $\beta$, passes through the ordered 
points ${\bf x}^{(\ell)}$ at times $\tau_\ell$ (see Fig. \ref{fig_multi_time}). 

To check the normalization of the above formula let us recall that 
\be
\int d{\bf y} P^{(N)}_{\rm OU}({\bf x},\tau|{\bf y}, \tau')
P^{(N)}_{\rm OU}({\bf y},\tau'|{\bf z}, \tau'')
=  P^{(N)}_{\rm OU}({\bf x},\tau|{\bf z}, \tau'') \label{95}
\ee 
where the integration $\int d{\bf y}$ here can either be over the ordered
coordinates, or unordered coordinates (it does not change the result,
since $P^{(N)}_{\rm OU}({\bf x},\tau|{\bf y}, \tau')$
vanishes unless the orders of ${\bf x}$ and ${\bf y}$ coincide). 
Integrating Eq. \eqref{94} on both sides over all variables except ${\bf x}^{(1)}$
it is easy to see, using the symmetry of $\tilde P_\beta$ under any permutation
of each set $(x_1^{(\ell)},\cdots x_N^{(\ell)})$ for each $\ell$, and \eqref{95},
that one recovers formula (\ref{PTxN3}).

Hence we have shown that the finite temperature fermion JPDF 
$\tilde P_{\beta}({\bf x}^{(1)},\cdots {\bf x}^{(m)})$ can be identified with
the multi-time JPDF of the positions of $N$ non-intersecting OU process
wrapped on a cylinder of perimeter $\beta$. This identification
demonstrates that for any fixed $N$ there is a classical interpretation 
as a stochastic process behind the finite temperature fermions. 
Note however that for any finite $N$ (in the canonical ensemble) the
fermion positions do not form a determinantal point process, except at
zero temperature. Consequently, the non-intersecting OU process 
also is not determinantal for fixed $N$, except in the zero temperature limit
($\beta \to +\infty$ limit). However, as we have discussed in previous sections, 
in the grand canonical ensemble
the finite temperature fermions form an extended determinantal process, thanks
to the EM theorem. For large $N$, using the equivalence between canonical and
grand canonical ensembles it then follows that all multi-time correlations of the
time-periodic OU process also become determinantal in the limit $N \to +\infty$, 
at any finite temperature. 

\subsection{Periodic Airy process} 

We can now define the periodic version of the Airy$_2$ process as follows.
It can be done equivalently on the finite temperature fermions and on the non-intersecting time-periodic
OU processes, using the above equivalence. For convenience we start with the
fermion picture. We are interested in the trajectory in imaginary time $\tau$ of the rightmost fermion at 
finite temperature. Using the equivalence to the OU process, this corresponds to the
top of the $N$ non-intersecting paths, propagating over the cylinder of perimeter $\beta$. 
At zero temperature, when the cylinder is of infinite perimeter, this process, properly centered and scaled in the large $N$ limit, is precisely the Airy$_2$ process. The finite perimeter of the cylinder $\beta$ deforms this process which we call the ``periodic Airy$_2$ process''. 

In order to describe this process more precisely, in principle we should start with the 
formula for $\tilde P_{\beta}({\bf x}^{(1)},\cdots, {\bf x}^{(m)})$ in \eqref{Pbeta}
which describes the JPDF of the positions of the $N$ fermions at $m$ different times
$\tau_1<\cdots <\tau_m$. Ordering the coordinates at each time slice and defining
the rightmost position at time $\tau_k$ as 
$x_{\rm max}^{(\ell)} = \max_{1 \leq i \leq N} x_i^{(\ell)}$, we would
like to integrate over the lowest $N-1$ positions at each time slice. This will give us
the periodic Airy$_2$ process which is characterized by the multi-time joint cumulative
distribution function (JCDF)
\be
{\rm Prob}( x_{\rm max}^{(1)}  < z_1 , \cdots ,   x_{\rm max}^{(m)}  < z_m) \;.
\ee
Using the equivalence to OU processes this JCDF then also describes the
multi-time CDF of the position of the top path. Note that by construction this 
process is periodic, of period $\beta$, in the time direction. 

In the large $N$ limit the position of the rightmost fermion is described
by the edge statistics. Hence we consider the rescaled process $\xi(u)$ in
rescaled time
\bea
&& \xi^{(k)} = \xi(u_k)= \frac{ x_{\rm max}^{(1)} - x_{\rm edge}}{w_N} \\
&& u_k = \omega \tau_k N^{1/3} \;,
\eea 
where we recall that $x_{\rm edge} = \sqrt{2\, N}\sqrt{\omega}$ and $w_N = N^{-1/6}/\sqrt{2 \omega}$ as in Eq. (\ref{wN}). Note that the time scale involved in this process is $\tau \sim w_N^2 \sim N^{-1/3}$,
a manifestation of Brownian scaling.
For large $N$ it converges to an extended determinantal process described
by the finite temperature extended Airy kernel given in Eqs. \eqref{scaledge}-\eqref{TextAiry0}. %\eqref{TextAiry1}. 
We will denote this process as the periodic
Airy$_2$ process
\be
\xi(u) \equiv {\cal A}_2^b(u) \;,
\ee
which depends on the single dimensionless parameter $b = \beta \omega N^{1/3}$.
By construction this process is periodic in time with period $b$. 
It follows from its definition that the multi-time JCDF of the rescaled process ${\cal A}_2^b(u)$ 
is given by a Fredholm determinant with the extended finite temperature
kernel (\ref{TextAiry1}). 

To be more precise we start with the one-time CDF $F_2^b(s)$ of the periodic Airy$_2$ process
\be
F_2^b(s) := {\rm Prob}( {\cal A}^b_{2}(u) < s )  = {\rm Det}[ I - P_s {\cal K}^{\rm edge}_b P_s]  \label{Airy1pt} 
\ee
where $P_s$ is the projector on the interval $[s,+\infty[$, with the edge kernel 
\be
{\cal K}_b^{\rm edge}(s,s') = \int_{-\infty}^{+\infty} dv \frac{1}{ e^{- b v} +1 } \Ai(s+v) \Ai(s'+v) \;,
\ee
where ${\cal K}_b^{\rm edge}(s,s') \equiv {\cal K}_b^{\rm edge}(s,s',u=0)$ is obtained
from \eqref{TextAiry0} by setting $u=0$. The Fredholm determinant $F_2^b(s)$  \eqref{Airy1pt}  
based on the kernel ${\cal K}_b^{\rm edge}(s,s')$
converges to the Tracy Widom GUE distribution, $F_2(s)$, in the limit $b \to +\infty$. 
For any finite $b$ it gives the centered and scaled distribution of the position $x_{\rm max}(T)$ of the
rightmost fermion at finite temperature and has also been found \cite{DLMS15,DLMS16} 
to identify with a generating function which appears in the
KPZ equation growth problem at finite time, with the correspondence $t_{\rm KPZ}=b^3$. 
Note that $F_2^b(s)$ here is called $Q_b(s)$ in Eq. (137) of \cite{DLMS16} 
and the scaled edge density, ${\cal K}_b^{\rm edge}(s,s)$, is denoted by $F_{1,b}(s)$ 
in Eq. (126) of \cite{DLMS16}.

Let us consider now the two-time JCDF of the periodic Airy$_2$ process. It is expressed as
the Fredholm determinant of a $2 \times 2$ matrix kernel 
\be
{\rm Prob}( {\cal A}^b_{2}(u_1) < s_1, {\cal A}^b_{2}(u_2) < s_2) 
= {\rm Det}
\begin{pmatrix} I - P_{s_1} K_0 P_{s_1} & -  P_{s_1} K_{u_1-u_2} P_{s_2}   \\  
 - P_{s_2} K_{u_2-u_1} P_{s_1}  & I - P_{s_2} K_0 P_{s_2} \end{pmatrix}
 \label{Airy2pt} 
\ee 
where we denote $K_{u,u'}=K_{u-u'}$ with the definition
\bea
%&& K_0(s,s') = {\cal K}_b^{\rm edge}(s,s',u=0) =  \int_{-\infty}^{+\infty} dv \frac{1}{ e^{- b v} +1 } \Ai(s+v) \Ai(s'+v) \\
K_u(s,s') :=  {\cal K}_b^{\rm edge}(s,s',u) &=& \begin{cases}& \int_{-\infty}^{+\infty} dv \frac{e^{- u v}}{ e^{- b v} +1 } \Ai(s+v) \Ai(s'+v) \quad , \quad 0 \leq u < b \label{TextAiry0b}  \\
& \\
 & - \int_{-\infty}^{+\infty} dv \frac{e^{- u v}}{e^{b v} +1} \Ai(s+v) \Ai(s'+v) \quad , \quad - b \leq u < 0 \end{cases} \label{TextAiry1b} 
\eea 
and we have $K_0(s,s') = {\cal K}_b^{\rm edge}(s,s')$. The kernel $K_{u}$ is antiperiodic
of period $b$ (see section \ref{sec:bulkedge}) which allows to obtain its value for arbitrary $u$ from the above equation.
The antiperiodicity of the kernel guarantees that all correlation functions of the process
are periodic in time. Let us note that in the zero temperature limit, i.e. for $b \to +\infty$ one recovers exactly the two-time JCDF of the standard Airy$_2$ process [see formula \eqref{Airy2pt0}].
This is more general, and extends to arbitrary multi-time correlations, i.e. one has ${\cal A}^b_{2}(u) \to {\cal A}_{2}(u)$ as ${b \to \infty}$, as a process in $u$. 

It is useful to define
\bea
K_{u;s_1,s_2}(s,s')= K_u(s+s_1,s'+s_2) = {\cal K}_b^{\rm edge}(s+s_1,s'+s_2,u) \;.
\eea 
Then one can rewrite, in more compact notations, the Fredholm determinant of
the $2 \times 2$ matrix kernel
\bea
{\rm Prob}( {\cal A}^b_{2}(u_1) < s_1, {\cal A}^b_{2}(u_2) < s_2) \nonumber
&=& {\rm Det}
\begin{pmatrix} I - P_0 K_{0;s_1,s_1} P_{0} & -  P_{0} K_{u_1-u_2;s_1,s_2} P_{0}   \\  
 - P_{0} K_{u_2-u_1;s_2,s_1} P_{0}  & I - P_{0} K_{0;s_2,s_2} P_{0} \end{pmatrix}  \label{FD2} \\
&=& {\rm Det} \left[ \delta_{ij} I - P_0 K_{u_i-u_j;s_i,s_j} P_0 \right]_{1 \leq i,j \leq 2} \;.
\eea

\subsection{Application: tails of the two-point function} 
\label{sec:applications}

As an application we will extract the tails of the two-point function
for large $s_1,s_2$ and focus on two cases: zero temperature ($b$ large)
and high temperature ($b \ll 1$).
As will be discussed below, for these tails we can consider the  trace expansion
to second order of the Fredholm determinant in \eqref{FD2}. We denote the
block matrix inside the Fredholm determinant in \eqref{FD2} as $I - M$ 
where $M$ has the structure
\be
 M=\begin{pmatrix} A&B\\C&D \end{pmatrix} \label{M} \;.
\ee
We can now use the general expansion
\be
{\rm Det}(I - M) = 1 - {\rm Tr} \, M + \frac{1}{2} ( ({\rm Tr} \, M)^2 - {\rm Tr} \, M^2) + {\cal O}(M^3) .
\ee
In the following we will be interested in the limit where all $A,B,C,D$ are small
(in some sense to be made more precise below). For this it is useful to first substract the diagonal parts (setting
$B=C=0$) which leads to
\bea
  {\rm Det} \begin{pmatrix} I-A&-B\\-C&I-D \end{pmatrix} - 
{\rm Det} \begin{pmatrix} I-A&0\\0&I-D \end{pmatrix}  &=& 
{\rm Det} \begin{pmatrix} I-A&0\\0&I-D \end{pmatrix} 
\left( {\rm Det} \begin{pmatrix} I&-B (I-D)^{-1} \\-C (I-A)^{-1} &I \end{pmatrix}  - 1 \right) \nonumber  \\
&=&
{\rm Det}(I-A) {\rm Det}(I-D)  \left( {\rm Det}(I - B (I-D)^{-1} C (I-A)^{-1} ) - 1 \right) \nonumber
\\
& = & - {\rm Tr} B C + {\cal O}(B^2 C^2,BCA,BCD) \label{BC} \;.
\eea 
Using \eqref{BC} in \eqref{FD2}, and fixing $u_1 \geq u_2$ for convenience, we obtain, for large $s_1$ and $s_2$, the connected part of the two point joint CDF of the
${\cal A}_2^b$ process as
\bea
&& {\rm Prob}( {\cal A}^b_{2}(u_1) < s_1, {\cal A}^b_{2}(u_2) < s_2) 
- {\rm Prob}( {\cal A}^b_{2}(u_1) < s_1) {\rm Prob}( {\cal A}^b_{2}(u_2) < s_2)
\\
&& \simeq - {\rm Tr} [ P_{s_1} K_{u_1-u_2} P_{s_2}  K_{u_2-u_1} ] 
%= - {\rm Tr} [ \bar K_{u_1-u_2;s_1,s_2} \bar K_{u_2-u_1;s_2,s_1} ] 
 = - \int_{s_1}^{+\infty} ds \int_{s_2}^{+\infty} ds' K_{u_1-u_2}(s,s')  K_{u_2-u_1}(s',s)  \;. \label{previous}
%&& = \int dv dv' K_\Ai(s_1+v,s_1+v') K_\Ai(s_2+v,s_2+v') \frac{e^{- (u_1-u_2) (v-v')}}{(e^{- b v} + 1)(e^{b v'}+1)} \\
%&& =  \int dv dv' K_\Ai(s_1+v,s_1-v') K_\Ai(s_2+v,s_2-v') \frac{e^{- (u_1-u_2) (v+v')}}{(e^{- b v} + 1)(e^{-b v'}+1)}
\eea 
Using the definition of the finite temperature extension Tracy-Widom CDF $F^b_2(s)$ in \eqref{Airy1pt0}, and
taking derivatives of \eqref{previous} with respect to $s_1$ and $s_2$, we obtain the JPDF 
$P(s_1,u_1;s_2,u_2) = \partial_{s_1} \partial_{s_2} {\rm Prob}( {\cal A}^b_{2}(u_1) < s_1, {\cal A}^b_{2}(u_2) < s_2)$ 
for $0 \leq u_2 < u_1 < \beta$ as
\bea
P(s_1,u_1;s_2,u_2) - \partial_{s_1} F^{b}_2(s_1) \partial_{s_2} F^{b}_2(s_2)  &=& - K_{u_1-u_2}(s_1,s_2) K_{u_2-u_1}(s_1,s_2) \\
&& \hspace{-3.5cm} =  \int_{-\infty}^{+\infty} dv \frac{e^{- (u_1-u_2) v}}{ (e^{- b v} +1) } \Ai(s_1+v) \Ai(s_2+v) \times 
\int_{-\infty}^{+\infty} dv' \frac{e^{- (u_2-u_1) v'}}{e^{b v'} +1} \Ai(s_1+v') \Ai(s_2+v') \label{AiAi} \\
&& \hspace{-3.5cm} =\tilde \phi_b(s_1,s_2,u_1-u_2)  \phi_b(s_1,s_2,u_1-u_2) \;,
\eea 
where we have defined, for future convenience, the two integrals depending on the variable $u$ 
\bea
&& \phi_b(s_1,s_2,u) = \int_{-\infty}^{+\infty} dv \frac{e^{u v}}{e^{b v} +1} \Ai(s_1+v) \Ai(s_2+v) \\
&& \tilde \phi_b(s_1,s_2,u) = \int_{-\infty}^{+\infty} dv \frac{e^{- u v}}{e^{- b v} +1} \Ai(s_1+v) \Ai(s_2+v) \;. \label{defphi} 
\eea
They satisfy the relation 
\be \tilde \phi_b(s_1,s_2,u) =
\phi_b(s_1,s_2,b -u)  \;.
\ee
Note that for $b=0$ it reduces to the Airy propagator, provided $u>0$ (or, more generally for $u$ complex with a non-negative real part) 
\be
2 \phi_{b=0}(s_1,s_2,u) = 2 \tilde \phi_{b=0}(s_1,s_2,-u) = 
\int_{-\infty}^{+\infty} dv e^{u v}  \Ai(s_1+v) \Ai(s_2+v) =
\frac{1}{\sqrt{4 \pi u}} 
e^{- \frac{(s_1-s_2)^2}{4 u} - \frac{1}{2} u (s_1+s_2) + \frac{u^3}{12} } \;. \label{AiryP}
\ee
We now use these relations to study the various interesting limits $b \to \infty$ and $b \to 0$.\\

{\it (i) Zero temperature limit $b \to +\infty$.}

At zero temperature, changing $v' \to - v'$ in the second integral in \eqref{AiAi} and taking $b \to +\infty$ limit gives
\bea
P(s_1,u_1;s_2,u_2) - \partial_{s_1} F_2(s_1) \partial_{s_2} F_2(s_2)  \\
&& \hspace{-3cm} =  \int_{0}^{+\infty} dv e^{- (u_1-u_2) v}  \Ai(s_1+v) \Ai(s_2+v) \times 
\int^{+\infty}_{0} dv' e^{- (u_1-u_2) v'} \Ai(s_1-v') \Ai(s_2-v') \;. \nonumber 
\eea
We now give only the large $u_1-u_2$ behavior
\be
P(s_1,u_1;s_2,u_2) - \partial_{s_1} F_2(s_1) \partial_{s_2} F_2(s_2) \approx  \frac{\Ai(s_1)^2 \Ai(s_2)^2}{|u_1-u_2|^2} \;, \label{120}
\ee
which is consistent, in the limit where $s_1,s_2$ are large, with the known result valid for any $s_1,s_2$
\cite{PraSpo02,AM04} 
\bea
{\rm Prob}( {\cal A}_{2}(u_1) < s_1, {\cal A}_{2}(u_2) < s_2) = F_2(s_1) F_2(s_2) + \frac{F_2'(s_1) F_2'(s_2)}{|u_1-u_2|^2} + {\cal O}(|u_1-u_2|^{-4})  \;.
\eea 
Note that the coefficient of the $1/|u_1-u_2|^2$ term can be retrieved by keeping all orders in $A,D$ in the second
line of Eq.~\eqref{BC}. Indeed taking a derivative of \eqref{Airy1pt} w.r.t. $s$ leads to the following identity
$F_2'(s) = F_2(s) \langle Ai | (1- P_s K_{\Ai} P_s)^{-1} | Ai \rangle$. Noting that at large $|u_1-u_2|$, 
$B$ and $-C$ become projectors equal to $| Ai \rangle \langle Ai|/|u_1-u_2|$, one obtains that  $- F_2(s_1) F_2(s_2) {\rm Tr} B (I-D)^{-1} C (I-A)^{-1} \simeq \frac{F_2'(s_1) F_2'(s_2)}{|u_1-u_2|^2}$ 
at large $|u_1-u_2|$. \\

{\it (ii) High temperature limit $b \to 0$.}

To investigate the small $b$ regime,
it is first convenient to rewrite \eqref{AiAi} in the form
\bea
P(s_1,u_1;s_2,u_2) - F_2'(s_1) F_2'(s_2)  
= \phi_b(s_1,s_2,b -( u_1-u_2)) \phi_b(s_1,s_2,u_1-u_2) \label{prod}
\eea 
where we have defined the integrals in \eqref{TextAiry1}.
Although one already see in formula \eqref{AiryP} the diffusion kernel appearing, as it will also appear below, it does not
help us here to evaluate \eqref{prod} which is restricted to the interval $0 \leq u_2 < u_1 <b$.
To analyze \eqref{prod} at small $b$, since the dependence in $u_1,u_2$ is periodic with period $b$ it is natural to rescale
\bea
u_1 = b \tilde u_1 \quad , \quad u_2 = b \tilde u_2
\eea 
where $0 \leq \tilde u_1 \leq \tilde u_2 < 1$. Let us first recall that the one-point CDF 
$F_2^b(s)$ of the process, in the limit $b \ll 1$ and $s \gg 1/b$
is given by \cite{DLMS16}
\bea
\partial_s F_2^b(s) \simeq \frac{1}{\sqrt{4 \pi b}} e^{-b s} \label{onepointT}  \;.
\eea
This shows that in the small $b$ limit the natural rescaling of $s$ is 
\be
s_1 = \tilde s_1/b \quad , \quad s_2 = \tilde s_2/b
\ee
and we study $\tilde s_1, \tilde s_2$ of order one, but large, which is the tail regime.
In addition, since $u_1-u_2 \sim b$ the Brownian scaling of the underlying OU process indicates that one
should focus on the regime $s_2-s_1 \sim \sqrt{b}$, hence we denote $s_2-s_1= \hat s_{21} \sqrt{b}$. 
One can thus rewrite \eqref{defphi} as
\bea
\phi_b(s_1,s_2,u_1-u_2) &=& \int_{-\infty}^{+\infty} dv \frac{e^{b (\tilde u_1- \tilde u_2) v}}{e^{b v} +1} \Ai(s_1+v) \Ai(s_1 + \hat s_{21} \sqrt{b} +v) \\
&=& \frac{1}{b}  \int_{-\infty}^{+\infty} dw \frac{e^{ (\tilde u_1- \tilde u_2) w}}{e^{w} +1} \Ai\left(\frac{\tilde s_1+w}{b}\right) 
\Ai\left(\frac{\tilde s_1 + w}{b} + \hat s_{21} \sqrt{b}\right) \;, \label{phi2}
\eea
where we have defined $w= b v$. This integral is dominated by the region $w + \tilde s_1<0$. In this regime we
use the fact that the product of Airy functions in the integral can be replaced by
\bea
{\rm Ai}((w+\tilde s_1)/b) {\rm Ai}((w+\tilde s_1)/b + \sqrt{b} \hat s_{21} ) \to
\frac{\sqrt{b} }{2 \pi \sqrt{|\tilde s_1+w|} } \cos( \sqrt{|\tilde s_1+w|} \hat s_{21}) 
\eea 
which amounts to neglect fast oscillating terms. Inserting in \eqref{phi2} and making a change of variable
$y=\sqrt{|\tilde s_1+w|}$ we obtain
\bea
&& \phi(s_1,s_2,u_1-u_2) = \frac{1}{\pi \sqrt{b}} 
e^{- (\tilde u_1-\tilde u_2) \tilde s_1} \int_0^{+\infty} dy \cos( y \hat s_{21} ) e^{- (\tilde u_1 -\tilde u_2) y^2} 
=  \frac{1}{\sqrt{4 \pi b (\tilde u_1-\tilde u_2)}}  e^{- \frac{\hat s_{21}^2}{4 (\tilde u_1-\tilde u_2)} - (\tilde u_1-\tilde u_2) \tilde s_1 }
\eea
where we have replaced $1/(1+ e^{- y^2 - \tilde s_1} ) \simeq 1$, since we study the tail $\tilde s_1 \gg 1$. 

\bea
&& P(s_1,u_1;s_2,u_2) - \frac{1}{4 \pi b} e^{- 2 \tilde s_1} = \frac{1}{4 \pi b } e^{- \tilde s_1} \frac{1}{\sqrt{(\tilde u_1-\tilde u_2) (1- (\tilde u_1-\tilde u_2))}}  
 e^{- \frac{\hat s_{21}^2}{4 (\tilde u_1-\tilde u_2)} - \frac{\hat s_{21}^2}{4 (1-(\tilde u_1-\tilde u_2))}  } + {\cal O}(e^{-2 \tilde s_1}) \;. \label{Pfinal}
\eea
This is the finite temperature analog of the zero temperature result given in \eqref{120}.
Note that in the limit $\tilde u_1-\tilde u_2 \to 0^+$ we find
\bea
&& P(s_1,u_1;s_2,u_2) \underset{\tilde u_1-\tilde u_2 \to 0^+}{\longrightarrow} \frac{1}{\sqrt{4 \pi b}} e^{-b s}  \delta(s_1-s_2) 
\simeq \partial_{s_1} F^b(s_1) \delta(s_1-s_2) \;,
\eea
which is consistent since one must recover the one-point PDF in that limit. 
In the picture of the OU process, the result in \eqref{Pfinal} has the following nice interpretation in the original variables $s,u$.
Consider the top path of the periodic OU process. Given its position $s_1$ at time $u_1$, 
the probability to propagate to a position $s_2$ at time $u_2$ is simply given, for large values of $s_1$, by the propagator of a Brownian bridge on the interval $[0,b]$. This reflects the fact that in the high temperature limit and for high values of $s_1,s_2$, the non-crossing condition becomes irrelevant in the dynamics. Note that the one point function 
\eqref{onepointT} is different from the one of the single particle OU process which is a Gaussian~\eqref{statbeta}.

\bigskip

\section{Universality of the time-periodic Airy$_2$ process}
\label{sec:univ} 

Until now we have studied specifically the example of the Ornstein-Uhlenbeck process which corresponds to
the quantum harmonic oscillator. Let us start with the single particle problem. One can consider a larger class of stochastic processes described by the Langevin equation of a particle diffusing in a classical potential $U(x)$
\be
\frac{d x(\tau)}{d \tau} = - U'(x(\tau)) + \eta(\tau)  \label{OUG1} 
\ee
where $\eta(\tau)$ is a centered Gaussian white noise, with 
correlator $\overline{ \eta(\tau) \eta(\tau')}=\delta(\tau-\tau')$.
For $U(x)=\frac{1}{2} \mu_0 x^2$ this corresponds to the OU process in \eqref{OU1}.
For general $U(x)$ the propagator for this classical process is given by \cite{Risken}
\be
P_U(x,\tau|x_0,\tau_0) = e^{- U(x)} G_V(x,\tau|x_0,\tau_0) e^{U(x_0)} \label{propG} 
\ee
where $G_V(x,\tau|x_0,\tau_0)= \langle x| e^{- H (\tau-\tau_0)} | x_0 \rangle$ is the quantum propagator 
associated to the Hamiltonian
\be
H = - \frac{1}{2} \partial_x^2 + V(x) \quad , \quad V(x) = \frac{1}{2} (- U''(x) + U'(x)^2)  \;, \label{HH} 
\ee
which is such that the ground state energy of $H$ is automatically zero. It is easy to see that the ground state wave function $\psi_0(x)$ is given by, up to a normalization constant 
$\psi_0(x) \sim e^{- U(x)}$ and the stationary measure of the process \eqref{OUG1} is given by 
$P_{\rm stat}(x) \sim e^{-2 U(x)}$ which is assumed to be normalizable. In the case of the OU process one has $V(x)=\frac{1}{2} \mu_0 x^2 - \frac{1}{2}$ [see Eq. \eqref{osc}].

We now consider $N$- non intersecting particles, each evolving with the Langevin equation \eqref{OUG1}. Following the OU example, as described in section \ref{sec:Nfermions}, this $N$ non-crossing particle process can be mapped onto the quantum problem of $N$ non interacting fermions trapped in the quantum potential 
$V(x)$. We can now apply the universality properties which were shown in \cite{DLMS16} and that we now recall. We assume that the quantum potential $V(x)$ is smooth and sufficiently confining. In this case
the average density at zero temperature is given by
\be
\rho_N(x) = \frac{1}{N} K_N(x,x) = \frac{1}{\pi N} \sqrt{ 2(\mu - V(x))} \theta(\mu- V(x)) \label{rhogen} 
\ee 
where $\theta(z)$ is the Heaviside theta function (i.e., $\theta(z) = 1$ if $z \geq 0$ and $\theta(z) = 0$ if $z<0$) and $\mu$ is determined by the normalization condition $\int dx \rho_N(x)=1$. This density has an edge 
at $x=x_{\rm edge}$ where the density is vanishing, i.e.,   
\be
V(x_{\rm edge}) = \mu \;.
\ee
In \cite{DLMS16} it was shown that at finite temperature the one-time correlation functions are given, 
in the edge region, by determinants of a kernel which takes the same scaling forms
as for the harmonic oscillator with renormalized width and reduced temperature (see equations (281) and (282) 
in \cite{DLMS16})
\be
w_N = (2 |V'(x_{\rm edge})|)^{-1/3} \quad , \quad b = 2^{-1/3} |V'(x_{\rm edge})|^{2/3} \beta  \;. \label{wNN}
\ee
It is now easy to see that the multi-time correlations for the more general process described
above are again given by determinants of an extended kernel. This extended kernel takes a universal form at the edge given by
\bea
K(x,\tau_i; y,\tau_j;\beta,\mu)  \simeq \frac{1}{w_N} {\cal K}^{\rm edge}_b\left(\frac{x-x_{\rm edge}}{w_N}, 
\frac{y-x_{\rm edge}}{w_N}, (\tau_i - \tau_j)b/\beta\right) \label{scaledgegen} 
\eea 
where ${\cal K}^{\rm edge}_b$ is the same function \eqref{TextAiry1} as for the harmonic oscillator. The scale factors $w_N$ and $b$ depend on the explicit form of $V(x)$ and are given in \eqref{wNN}. 

In conclusion this shows that the top path of $N$ non-crossing particles diffusing in the generic potential $U(x)$ as in \eqref{OUG1}, wrapped around a cylinder of perimeter $\beta$, properly centered and scaled, is described by the universal periodic Airy$_2$ process. This holds provided the corresponding quantum potential is sufficiently confining to provide the soft-edge universality of the fermion problem. Equivalently this universality also holds for the multi-time (in imaginary time) correlation functions for the trapped fermions.

\section{Application to real time equilibrium dynamics of trapped fermions: dynamical density-density correlations} 
\label{sec:real}

Here we consider the real time quantum dynamics of fermions. We restrict to {\it equilibrium dynamics} at finite temperature $T$, which can be obtained from the analysis of the previous sections.

\subsection{Real time quantum dynamics of a single particle}

Let us recall first the one-particle problem. Let us recall that in quantum mechanics the time dependent average (over measurements) denoted $\langle O \rangle_t$ of an observable $\hat O$ (for instance the position $\hat x$ or the momentum $\hat p$ etc..), is given by the expectation value
\be
\langle O \rangle_{\psi(t)} = \langle \psi(t) | \hat O | \psi(t) \rangle = \langle \psi(0) | \hat O(t) | \psi(0) \rangle
= {\rm Tr} \left[ | \psi(0) \rangle \langle \psi(0)| \,\hat O(t) \right]
\ee
where $| \psi(t) \rangle$ is the quantum state of the system at time $t$. By definition
\be
\hat O(t) = e^{i \hat H t/\hbar} \, \hat O \, e^{-i \hat H t/\hbar}
\ee
is the observable in the Heisenberg representation (where it becomes time dependent),
which allows to take expectation values over the initial state $| \psi(0) \rangle$. Here $\hat H$ is the Hamiltonian of the system, and we restrict to time independent Hamiltonians $\hat H$ (in the time dependent case one needs time-ordered exponentials). If several measurements are performed successively, of observables $\hat O_j$ at times $t_j$, with $t_1<t_2<..<t_m$, then the joint average (over quantum measurements)
is given by the expectation value of the product
\be
\langle O_1(t_1) \cdots O_m(t_m) \rangle_{\psi(0)} =
\langle \psi(0) | \hat O_m(t_m) \cdots \hat O_1(t_1) | \psi(0) \rangle 
= {\rm Tr} \left[ | \psi(0) \rangle \langle \psi(0)| \, \hat O_m(t_m) \cdots \hat O_1(t_1) \right] \;, \label{corrpsi} 
\ee
where $\hat O_j(t_j) = e^{i \hat H t_j/\hbar} \, \hat O_j \, e^{-i \hat H t_j/\hbar}$.

At finite temperature, in the canonical ensemble, the system is described not by a single quantum state but by a statistical mixture of
states represented by a density matrix, and one must replace in the above formula
\be
| \psi(0) \rangle \langle \psi(0)| \to  \frac{1}{Z(\beta)} \sum_k |\phi_k \rangle e^{- \beta \epsilon_k } \langle \phi_k |
= \frac{1}{Z(\beta)} e^{- \beta \hat H} 
\ee
and obtain the {\it correlation function} at temperature $T=1/\beta$
\be
\langle O(t_1) \cdots O(t_m) \rangle_T \, = \frac{1}{Z(\beta)}  {\rm Tr} \left[ e^{-\beta \hat H} 
 \, \hat O_m(t_m) \cdots \hat O_1(t_1) \right] 
\ee
with $Z(\beta)= {\rm Tr} e^{- \beta \hat H}$.
It is immediate to see that the one-time observable $\langle O_1(t_1) \rangle_T = \frac{1}{Z(\beta)}  {\rm Tr} \left[ e^{-\beta \hat H} 
 \, \hat O(t_1) \right] = \frac{1}{Z(\beta)}  {\rm Tr} \left[ e^{-\beta \hat H}  \, \hat O \right]$ is independent on $t_1$ and given by the canonical equilibrium expectation value. 
 The $n$-time correlations with $m >1$ are however non-trivial functions of the $m-1$ time differences $t_2-t_1, \cdots, t_m-t_{m-1}$, a property called time translational invariance (TTI). These functions
 describe the {\it quantum equilibrium dynamics}. Note that if one considers instead evolution
from a pure state $|\psi(0) \rangle$ as above, the $m$-time correlations are in general 
not TTI (hence they depend on all the $m$ times) unless the initial state is an eigenstate of $\hat H$
in which case they are TTI (see example below). This is the case for instance for the ground state. 

By choosing in \eqref{corrpsi} the operators $\hat O_j = |x^{(j)} \rangle \langle x^{(j)}|$ 
and the initial condition $\psi(0)=\phi_k$, an eigenstate $k$ of the harmonic oscillator (for instance the ground state) we define a real-time quantum correlation function involving
times $t_j$ as
\bea
C_{\phi_k}(x^{(1)},\cdots , x^{(m)}) &=& e^{i \epsilon_k (t_m-t_1)}
\langle \phi_k | x^{(m)} \rangle \langle x^{(m)} | e^{- i \hat H (t_m-t_{m-1})} |x^{(n-1)} \rangle \cdots \langle x^{(2)} | e^{- i \hat H (t_2-t_1)} |x^{(1)} \rangle \langle x^{(1)} | \phi_k \rangle \label{136} \\
& =& e^{i \epsilon_k (t_m-t_1)} \langle \phi_k | x^{(m)} \rangle G(x^{(m)},x^{(m-1)};i (t_m- t_{m-1})) 
\cdots G(x^{(2)},x^{(1)};i (t_2-t_{1}))
 \langle x^{(1)} | \phi_k \rangle \,, \nonumber
\eea
where $G(x,x';\tau) \equiv G(x,\tau;x',0)$ is the Euclidean quantum mechanical propagator of the harmonic oscillator, given in Eq. \eqref{propdec} taken at argument $\tau \to i t$ since we are considering here the real time dynamics.

Similarly at finite temperature one defines the following quantum correlation at 
times $t_j$ 
\bea
C_{T}(x^{(1)},\cdots , x^{(m)}) &=& \frac{1}{Z(\beta)}  \langle x^{(1)} | e^{\hat H (i (t_m-t_1) - \beta)}| x^{(m)} \rangle \langle x^{(m)} | e^{- i \hat H (t_m-t_{m-1})} |x^{(m-1)} \rangle \cdots \langle x^{(2)} | e^{- i \hat H (t_2-t_1)} |x^{(1)} \rangle   \label{145} \\
& =& \frac{1}{Z(\beta)} G(x^{(1)}, x^{(m)}; \beta - i (t_m-t_1)) \rangle G(x^{(m)},x^{(m-1)};i (t_m- t_{m-1})) 
\cdots G(x^{(2)},x^{(1)};i (t_2-t_{1}))  \;. \nonumber
\eea

Let us now compare the expressions for the real time correlations in \eqref{136} and in \eqref{145} 
with the multi-time JPDF in imaginary time given in \eqref{PEE}, \eqref{norm1} and \eqref{Pbeta} 
respectively, upon setting $N=1$ for the single particle case that we are considering here. 
These results are related to each other by the replacement $\tau \to i t$ 
as expected \cite{FW71}. Indeed the expressions \eqref{PEE} and \eqref{Pbeta} can be reordered
using that the respective probabilities are ordered. Stationarity of the imaginary time process corresponds
to the TTI property of the real time evolution. There are however two important differences between these observables in real and imaginary times. First, the process is periodic of period $\beta$ in imaginary time, while it is defined on the real line for real time $t$.
Secondly, both real time correlations in \eqref{136} and \eqref{145} are not a priori real, hence
do not have a probabilistic interpretation, unlike their imaginary time counterparts.

\subsection{Real time quantum dynamics of non-interacting fermions in a harmonic trap}

\subsubsection{General framework} 

The real time dynamics of $N$ non-interacting fermions can be studied by similar methods. One substitutes
$|\psi(0) \rangle$ as a $N$-fermion state, $\hat H \to {\cal H}_N$, and $\hat O_j$ as $N$-body observables, in all formula above. Considering the operators $\hat O_j = | {\bf x}^{(j)} \rangle \langle {\bf x}^{(j)} |$ in \eqref{corrpsi}, Eq.
\eqref{136} generalises into a formula for the quantum correlation of the real time dynamics starting from an eigenstate $|E\rangle$
(pure state) of the $N$-body Hamiltonian. This formula is identical to the one for the quantum JPDF, 
$P_E({\bf x}^{(1)},\cdots, {\bf x}^{(m)})$, in Eq.~\eqref{PEE}
with the replacement $\tau_j \to i t_j$. Hence it has a determinantal property 
with exactly all the same formula for multi-time correlations given by determinants as in section \ref{sec:state}.
In particular if we choose $|E\rangle$ to be the ground state of the harmonic oscillator we obtain, replacing
$\tau \to i t$ in Eq. \eqref{KHext1}-\eqref{KHext2}, the real
time extended Hermite kernel 
\bea
&&K^r_N(x,t_i; y,t_j) = \sum_{k=0}^{N-1} e^{- i k \omega (t_j - t_i)} \phi^*_k(x) \phi_k(y) \; \; \;, \; t_i \geq t_j 
\label{KHext1t}
\\
&&K^r_N(x,t_i; y,t_j) = - \sum_{k=N}^\infty e^{- i k \omega (t_j-t_i) }\phi^*_k(x) \phi_k(y) \; \; \;, \; t_i < t_j \;,
\label{KHext2t} 
\eea
where the superscript $r$ indicates real time.
For instance we obtain the multi-time quantum correlation of the number density operator
\bea
\langle \hat \rho(y_1,t_1) \cdots \hat \rho(y_m,t_m) \rangle_{E} = \det_{1 \leq \ell,\ell' \leq m} 
K_N(y_{\ell} , t_\ell ; y_{\ell' }, t_{\ell'})  \;,
\eea
where the number density operator (i.e. normalized to $N$) is defined as
\be
\hat \rho(y,t) = e^{i \hat H t} \hat \rho(y) e^{- i \hat H t} \quad , \quad
\langle {\bf x}| \hat \rho(y) | {\bf x}' \rangle =  \sum_{i=1}^N \delta(x_i-y) \delta({\bf x}-{\bf x}') \;.
\ee

We can also study the real time dynamics of the fermion system at finite temperature. As in the case of imaginary time it is more convenient to consider the grand canonical ensemble. One finds again that the correlation functions can be written as determinants with the extended, grand canonical kernel in real time
\begin{eqnarray}
&&K^r(x,t_i; y,t_j ;\beta,\mu) = \begin{cases} & \sum_{k=0}^\infty \frac{e^{i (\epsilon_k-\mu)(t_i - t_j)} }{e^{\beta (\epsilon_k - \mu)} +1 }  \,  \phi^*_k(x) \phi_k(y) \; \; \;, \; t_i \geq t_j 
\label{Kext1t}
\\
& \\
&
%K(\tau_i,x; \tau_j,y; \beta,\mu) = 
- \sum_{k=0}^\infty 
\frac{e^{i (\epsilon_k-\mu)(t_i - t_j)} }{e^{-\beta (\epsilon_k - \mu)} +1 }  
\phi^*_k(x) \phi_k(y) \; \; \;, \, t_i < t_j \end{cases} 
\label{Kext2t} 
\end{eqnarray}
where for convenience we have inserted in the kernel a global factor $e^{- i \mu (t_i - t_j)}$ which cancels
out in any correlation function. This extended kernel is a function only of the time difference
$t_i-t_j$ and is not continuous at $t_i=t_j$ where it has a $\delta(x-y)$ jump
as in \eqref{jump}.
Note that the second line in \eqref{Kext2t} is obtained from the first one
by replacing $t_i - t_j \to - i \beta + t_i - t_j$. In fact this extended kernel is an anti-periodic in each time variable with
period $- i \beta$. 

As we have stated it before, in absence of the confining potential, i.e. when the system is
translationally invariant (free fermions), this kernel 
has been well studied in the solid-state physics literature. However there it is usually
presented in the frequency domain. To make contact with this literature, it is useful to give
the expression of this extended kernel in the general case in the frequency domain. We first reexpress
the kernel in \eqref{Kext1t} to make apparent the time translational invariance as
$K^r(x,t_i; y,t_j ;\beta,\mu) = K^r(x,y;t_i-t_j;\beta,\mu)$. One finds
\bea
&& \int_{-\infty}^{+\infty} dt  e^{-i \omega t - \eta |t|} K^r(x,y;t;\beta,\mu) = - i \sum_{k=0}^\infty \,  \phi^*_k(x) \phi_k(y) \left( \frac{1- \langle n_k \rangle}{\omega+ \mu - \epsilon_k + i \eta} 
+ \frac{\langle n_k \rangle}{\omega + \mu- \epsilon_k  - i \eta} \right) \;, 
\eea
where $\langle n_k \rangle$ is the Fermi factor given also in \eqref{fermi_factor}. 
%&& \bar \ran_k = \frac{1}{e^{\beta (\epsilon_k - \mu)} +1 } 

\subsubsection{Real time dynamics at the edge}  

As an application let us consider the real time dynamics near the edge. We obtain that it is
described by the real time extended edge kernel
\bea
K^r(x,t_i;y,t_j;\beta,\mu)  \simeq \frac{1}{w_N} {\cal K}^{\rm edge,r}_b\left(\frac{x-x_{\rm edge}}{w_N}, 
\frac{y-x_{\rm edge}}{w_N}, (t_i - t_j) \omega N^{1/3}\right) \label{scaledget} 
\eea 
in terms of the finite temperature real time extended Airy kernel 
\bea
&& {\cal K}_b^{\rm edge,r}(s,s',u) = \begin{cases} & \int_{-\infty}^{+\infty} dv \frac{e^{- i u v}}{ e^{- b v} +1 } \Ai(s+v) \Ai(s'+v)= \tilde \phi_b(s,s',i u) \quad , \quad u \geq 0 \label{TextAiry0t}  \\
& \\
&
% {\cal K}_b^{\rm edge}(s,s',u) =
 - \int_{-\infty}^{+\infty} dv \frac{e^{- i u v}}{e^{b v} +1} \Ai(s+v) \Ai(s'+v)  = - \phi_b(s,s',- i u)\quad , \quad u<0 \;,\end{cases}  \label{TextAiry1t} \;
\eea 
which is now defined for arbitrary real $u$. Note that these integrals are convergent.
Recalling that we have defined $\beta(\epsilon_k-\mu) = - b v$, 
it is easy to see that the first line in \eqref{TextAiry0t} corresponds to contributions from states below the Fermi surface and the second from states above the Fermi surface. 

From this kernel we can calculate the density-density correlation at temperature $T$ as
\bea
 \langle \hat \rho(x,t_1)  \hat \rho(y,t_2) \rangle_{T} &=&
\frac{1}{w_N^2}  g_b^{\rm edge}\left(\frac{x-x_{\rm edge}}{w_N},\frac{y-x_{\rm edge}}{w_N}, (t_1 - t_2) \omega N^{1/3}\right) \label{res1} \\
g_b^{\rm edge}(s,s',u)&=&{\cal K}_b^{\rm edge,r}(s,s,0) {\cal K}_b^{\rm edge,r}(s',s',0) - {\cal K}_b^{\rm edge,r}(s,s',u) {\cal K}_b^{\rm edge,r}(s,s',-u) \label{res2} 
\\
& = & \tilde \phi_b(s,s,0) \tilde \phi_b(s',s',0)  + \tilde \phi_b(s,s',i u) \phi_b(s,s',i u) \;, \label{res3}
\eea
which is a complex and even function of $u$, the last line being true only for $u>0$, and $b= \omega N^{1/3}/T$ and ${\cal K}_b^{\rm edge,r}(s,s',u)$
is given in \eqref{TextAiry1t}. Note that we can write the result in the frequency domain as
\bea
&& {\cal K}_b^{\rm edge,r}(s,s',\omega) = \int du e^{- i u \omega - \eta |u|} {\cal K}_b^{\rm edge,r}(s,s',u)
= - i \int_{-\infty}^{+\infty} dv \left[ \frac{\Phi_>(v)}{\omega+ v + i \eta} + \frac{\Phi_<(v)}{\omega+ v - i \eta} \right] \\
&& \Phi_>(v) = \frac{1}{e^{b v} +1} \Ai(s+v) \Ai(s'+v) \quad , \quad \Phi_<(v) = \frac{1}{e^{-b v} +1} \Ai(s+v) \Ai(s'+v) \;,
\eea
where $\eta=0^+$, and the functions $\Phi_<(v)$ and $\Phi_>(v)$ correspond to contributions from below and above the Fermi surface, respectively.

\subsubsection{Real time density-density correlation at the edge}

Let us now analyze the behavior of the density-density correlation function. For $u=0$ one recovers
the equal time equilibrium correlation $g_b^{\rm edge}(s,s',u=0)=g_b^{\rm edge}(s,s')$, given below Eq. (133) in \cite{DLMS16}, 
and which vanishes at coinciding points,
i.e., $g_b^{\rm edge}(s,s,u=0)=0$. For $u=0^+$ one has
\bea
g_b^{\rm edge}(s,s',u=0^+)= {\cal K}_b^{\rm edge,r}(s,s',0) \delta(s-s') \;.
\eea

In the large time difference limit $|t_2 - t_1| \to +\infty$ one expects that
\bea
 \langle \hat \rho(x,t_1)  \hat \rho(y,t_2) \rangle_{T} \to \langle \hat \rho(x,t_1) \rangle_T
 \langle \hat \rho(y,t_2) \rangle_T = \rho_N(x)  \rho_N(y) 
 \eea
where we have used the fact that the single time expectation value of the
density is time-independent. As shown in Eqs. (125)-(126) in \cite{DLMS16} 
it takes the scaling form near the edge 
\be
\rho_N(x) = \frac{1}{w_N}  F_{1,b}\left(\frac{x-x_{\rm edge}}{w_N}\right) \quad , \quad 
F_{1,b}(s) = {\cal K}_b^{\rm edge}(s,s) = {\cal K}_b^{\rm edge,r}(s,s,0) 
\ee
where ${\cal K}_b^{\rm edge,r}(s,s,u)$ is defined in \eqref{TextAiry1t}.
It can be seen indeed in \eqref{res2}-\eqref{res3} that the second term
decreases to zero as $u \to \infty$ leading to
\bea
\lim_{u \to \infty} g_b^{\rm edge}(s,s',u) = {\cal K}_b^{\rm edge,r}(s,s,0) {\cal K}_b^{\rm edge,r}(s',s',0) \;.
\eea
We thus define the connected part of the density-density correlation
\bea
 \langle \hat \rho(x,t_1)  \hat \rho(y,t_2) \rangle_{T}^c &=& \langle \hat \rho(x,t_1)  \hat \rho(y,t_2) \rangle_{T} 
 - \langle \hat \rho(x,t_1) \rangle_T \langle \hat \rho(y,t_2) \rangle_{T} \\
 & =&
\frac{1}{w_N^2}  g_b^{\rm edge,c}\left(\frac{x-x_{\rm edge}}{w_N},\frac{y-x_{\rm edge}}{w_N}, (t_1 - t_2) \omega N^{1/3}\right) \label{res1c} \\
g_b^{\rm edge,c}(s,s',u)&=& - {\cal K}_b^{\rm edge,r}(s,s',u) {\cal K}_b^{\rm edge,r}(s,s',-u) =
\tilde \phi_b(s,s',i u) \phi_b(s,s',i u) \;. \label{res3c}
\eea

We now study the low temperature scaling limit $b \to +\infty$ and $u \to +\infty$, keeping the ratio
$u/b$ fixed. 
We will use the following relation
\be
\tilde \phi_b(s,s',i u) +  \phi_b(s,s',i u) = \frac{1}{\sqrt{4 \pi i u}} 
e^{- \frac{(s-s')^2}{4 i u} - \frac{1}{2} i u (s+s') - i \frac{u^3}{12} } - 
2 i \int_{-\infty}^{+\infty} dv \frac{\sin(u v)}{e^{- b v} +1} \Ai(s+v) \Ai(s'+v) \label{sum} \;.
\ee
In this low temperature limit one easily shows that for $u>0$ and large $u \sim b \to +\infty$ one has
\bea
&& \int_{-\infty}^{+\infty} dv \frac{e^{- i u v}}{e^{- b v} +1} \Ai(s+v) \Ai(s'+v) 
= 
 \int_{-\infty}^{+\infty} dv \frac{i \sin(u v)}{e^{- b v} +1} \Ai(s+v) \Ai(s'+v) \simeq
- i \frac{\pi}{b \sinh(\frac{\pi u}{b})} \Ai(s) \Ai(s') \;. \label{asympt}
\eea
This implies
\bea
&& \tilde \phi_b(s,s',i u) \simeq - i \frac{\pi}{b \sinh(\frac{\pi u}{b})} \Ai(s) \Ai(s') \\
&& \phi_b(s,s',i u) \simeq \frac{1}{\sqrt{4 \pi i u}} 
e^{- \frac{(s-s')^2}{4 i u} - \frac{1}{2} i u (s+s') - i \frac{u^3}{12} } + {\cal O}\left(\frac{1}{b},\frac{1}{u}\right) 
\eea 
where we have used \eqref{asympt} and in \eqref{sum} we can neglect the last
term which is of order ${\cal O}(1/b)$ or ${\cal O}(1/u)$ as compared to the first term which is,
in modulus, of order ${\cal O}(1/\sqrt{u})$. Putting together these two results, the
connected part of the density-density correlation function
is given by (for $u>0$)
\bea\label{correl_asympt}
&& g_b^{\rm edge,c}(s,s',u) \simeq - \sqrt{ i \pi} \frac{\Ai(s) \Ai(s')}{2 b \sinh(\frac{\pi u}{b}) u^{1/2}}  
 e^{- i \left(- \frac{(s-s')^2}{4 u} + \frac{1}{2} u (s+s') +  \frac{u^3}{12} \right)} \;.
\eea
\begin{figure}
\includegraphics[width = 0.6\linewidth]{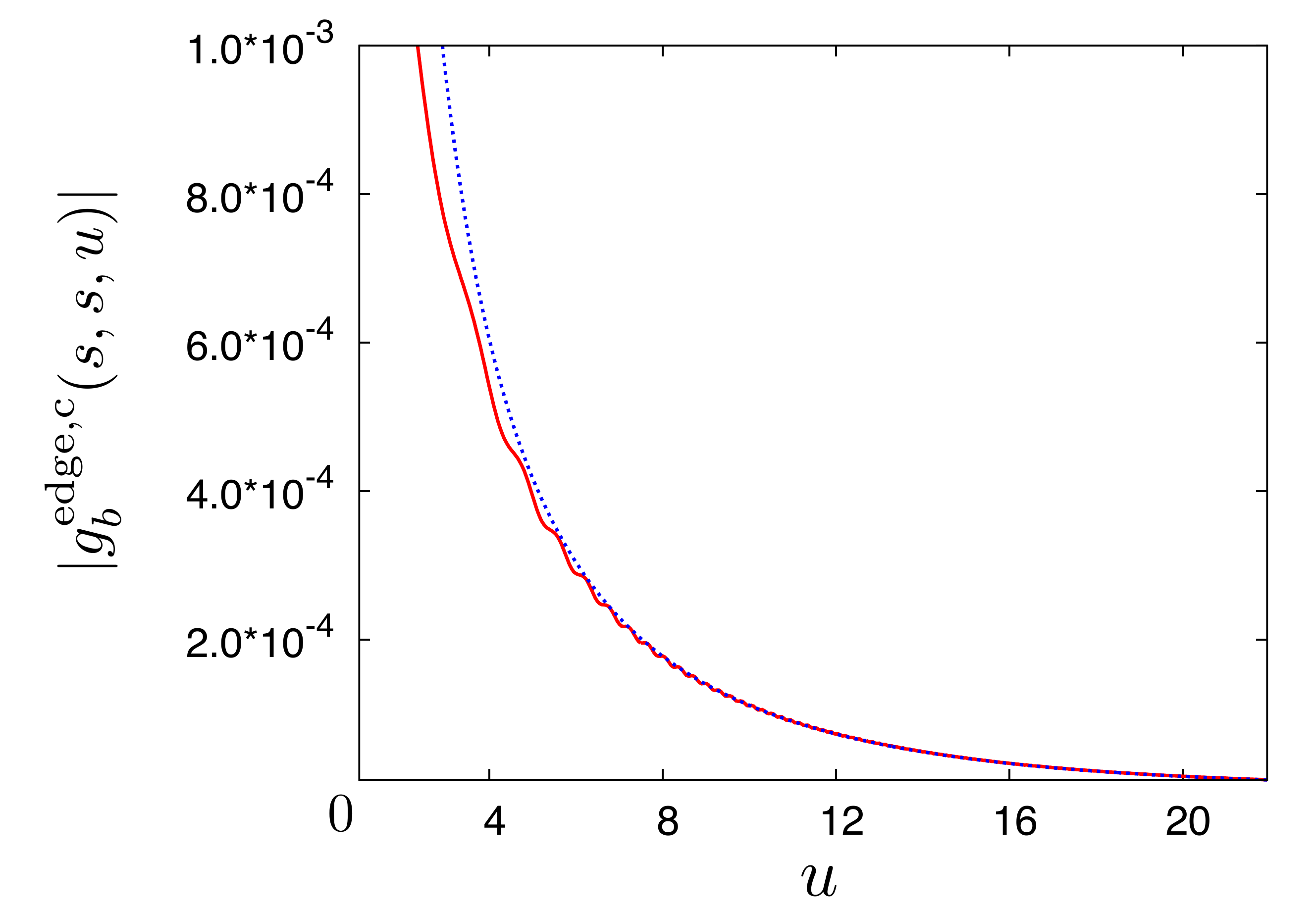}
\caption{Plot of the modulus of the connected correlation scaling function at coinciding points $g_{b}^{\rm edge,c}(s,s,u)$ for $s=1.0$ as a function of $u$ for $b=10.$. The solid line corresponds to the numerical evaluation of the exact formula given in Eq. (\ref{res3c}) while the dotted line corresponds to the asymptotic result (\ref{correl_asympt}) valid for large $u$ and large $b$ with $u \sim b$.}\label{fig_correl}
\end{figure}
We note that the last term is the quantum Airy propagator which describes a particle in a linear potential,
and we can interpret the Airy function factors as the amplitudes to create and destroy
a hole respectively at $s$ and $s'$ (which then propagates in a linear potential).
In the zero temperature limit $b \to +\infty$ the connected correlation scaling function
decays algebraically as a function of time as
\be
g_b^{\rm edge,c}(s,s',u) \simeq u^{-3/2} \;,
% - \sqrt{ i \pi} \frac{\Ai(s) \Ai(s')}{2 \pi u^{3/2}}  
% e^{- i \left(- \frac{(s-s')^2}{4 u} + \frac{1}{2} u (s+s') +  \frac{u^3}{12} \right)} 
\ee
while in the opposite limit $u \gg b$ it decays exponentially as
\be
g_b^{\rm edge,c}(s,s',u) \simeq u^{-1/2} \exp\left(- \frac{\pi u}{b}\right) \;.
% - \sqrt{ i \pi} \frac{\Ai(s) \Ai(s')}{2 \pi u^{3/2}}  
% e^{- i \left(- \frac{(s-s')^2}{4 u} + \frac{1}{2} u (s+s') +  \frac{u^3}{12} \right)} 
\ee
In Fig. \ref{fig_correl}, we show a plot of the modulus of $g_b^{\rm edge,c}(s,s,u)$ at coinciding point and as a function of $u$ and compare it with the asymptotic result in Eq. (\ref{correl_asympt}).

Let us compare this result at the edge with the result for free fermions
(non-interacting fermions in the absence of a confining potential),
at $T=0$ for simplicity \cite{GiamBook,Pereira}.
The real time Green's function (with the notations of the present paper)
\be
\int_{-\infty}^{+\infty} dx e^{- i k x} G^r_{\rm free}(x,t;0,0) = \left( - \theta(k-k_F) \theta(-t) + \theta(k_F-k) \theta(t)\right) e^{ i \frac{1}{2} k^2 t}
\ee 
where $k_F$ is the Fermi momentum.
The connected density-density correlation function is given by the product 
\be
\langle \hat \rho(x,t)  \hat \rho(0,0) \rangle_{T=0,{\rm free}}^c =
- G^r_{\rm free}(x,t;0,0) G^r_{\rm free}(x,-t;0,0) \;.
\ee
One of the factors is the hole Green's function (involving integration below the Fermi level) 
which at large time behaves~as 
\be
\int_0^{k_F} \frac{dk}{2 \pi} e^{- i k^2 t} \sim \int_0^{+\infty} \frac{dk}{2 \pi} e^{- i k^2 t} \sim t^{-1/2} \;,
\ee
while the other one is the particle Green's function (involving momentum integration above the Fermi level)
which decays at large time as
\be
\int_{k_F}^{\infty} \frac{dk}{2 \pi} e^{- i k^2 t}  \sim t^{-1} \;.
\ee
Note that the same results can also be derived {\it in the bulk}, in presence of a confining potential 
from Eq. \eqref{extendedsine0}. This is of course expected since deep inside the bulk the fermions do not feel the confining potential. Putting both factors together thus leads to an asymptotic $t^{-3/2}$ real time decay of the connected density correlation both for free fermions and in the bulk of confined fermions. Although the exponent $3/2$ is the same as the one obtained above in the case of the edge, the detailed mechanism seem different.
Indeed at the edge the $ t^{-1/2}$ decay comes from the particle contribution (above the Fermi surface) 
while the $t^{-1}$ decay comes from the hole contribution. Interestingly there seems to be a ``role reversal'', between the bulk and the edge, of the time decay.

\bigskip

\subsection{Joint measurements}

Let us finish this section on real time dynamics by mentioning the problem of 
joint quantum measurements at different times. For simplicity let us focus on 
two measurements at two different times ($m=2$). We start with a single particle and perform successive measurements of its
position at times $t_1$ and $t_2>t_1$. As is well known from the principles of quantum mechanics, 
the possible outcomes of the measurement of an observable $\hat O$ in the quantum state $|\psi \rangle$
are the eigenvalues
$\alpha$ of $\hat O$, obtained with respective probabilities $||P_\alpha |\psi \rangle||^2$
where $P_\alpha$ is the projector on the corresponding eigen-subspace ${\cal E}_\alpha$.
Importantly, just after the measurement the new quantum state is $P_\alpha |\psi \rangle|$.
Hence we must introduce projectors associated to the observable. If we measure position and
if the particle lives on a lattice the relevant projector is $P_x = |x \rangle \langle x|$. In the continuum
one may consider instead a projector e.g. on an interval $P_I = \int_{x \in I} dx |x \rangle \langle x|$
, which indeed statistfies $P_I^2=P_I$. Given these preliminaries, we can now write the
probability of the joint outcome of two successive measurements of the position as
\bea
&& {\rm Prob}(x_1 \in I_1, t_1;x_2 \in I_2, t_2) = {\rm Tr} \left[ P_{I_2} e^{- i \hat H (t_2-t_1)} P_{I_1} 
e^{- i \hat H t_1} \hat D e^{i \hat H t_1} P_{I_1} e^{i \hat H (t_2-t_1)} P_{I_2} \right]  \label{probmeasure}
\eea
where the last $P_{I_2}$ on the left can be omitted thanks to cyclicity of the trace.
In this formula $\hat D$ is the initial density matrix, that is $\hat D=|\psi(0) \rangle \langle \psi(0) |$ if one considers the evolution starting from a pure state, and
$\hat D = e^{- \beta \hat H}/Z(\beta)$ for the evolution in thermal equilibrium.
If we work on a lattice we can write the same formula, with
$P(x_1, t_1;x_2, t_2)$ on the l.h.s. (the probability to observe $x_1$ at $t_1$ and then $x_2$ at $t_2$)
and $P_{I_j}=|x_j \rangle \langle x_j|$, $j=1,2$.

The normalization of \eqref{probmeasure} is subtle. Clearly setting $I_1=I_2=\mathbb{R}$
yields unity, and setting either $P_{I_1}=1$ or $P_{I_2}=1$ yields the correct probabilities 
for a single measurement $P(x_j,t_j)=\langle x_j | e^{- i \hat H t_j} \hat D e^{i \hat H t_j}  |x_j \rangle$ 
%{\blue I do not understand what is the meaning of $I_1 = 1$ here, in the continuum setting. Can you clarify it a bit ?}. 
Hence setting e.g. $P_{I_1}=1$ means that no measurement
has been performed at time $t_1$. Going to the lattice case, we see that
$\sum_{x_2 \in \mathbb{Z}} P(x_1, t_1;x_2, t_2)=1$, but that 
$\sum_{x_1 \in \mathbb{Z}} P(x_1, t_1;x_2, t_2) \neq 1$. This is because
by performing a measurement at time $t_1$, one perturbs the system (one projects
its quantum state into one of the eigen-subspace associated to the observable)
and, even if one does not read the outcome of the measurement, it is different
from not performing the measurement. 

Consider now the evolution from an energy eigenstate $|E \rangle$. We first evolve this state for
a time $t_1$, perform the measurement at $t_1$, evolve further from time $t_1$ to $t_2$ and perform
the measurement at $t_2$. The probability of the outcome is then
\bea
&& {\rm Prob}(x_1 \in I_1, t_1;x_2 \in I_2, t_2) = 
\int_{I_2} dx_2 \, \bigg| \langle x_2 |e^{- i \hat H (t_2-t_1)} \int_{I_1} dx_1 |x_1 \rangle \langle x_1 |E \rangle\bigg|^2 \\
&& = \int_{I_1} dx_1 \int_{I_1} dx'_1 \int_{I_2} dx_2 \langle E |x_1 \rangle G(x_1,x_2;i (t_1-t_2)) G(x_2,x'_1;i (t_2-t_1))  
\langle x'_1 |E \rangle \label{prob1}
\eea
in terms of the quantum propagator 
\be \label{G_q} 
G(x,x_0;i t) = \langle x| e^{- i t  \hat H} |x_0 \rangle 
= \sum_{k=0}^{+\infty} \phi_k(x)  \phi_k^*(x_0) e^{- i  \epsilon_k t} \;.
\ee 

Let us now discuss the case of $N$ non-interacting fermions. The formula \eqref{prob1}
extends to this case, replacing $x_j \to {\bf x}^{(j)}$ the set of coordinates of the 
$N$ fermions, and $|E \rangle$ being the $N$-fermion energy eigenstate, 
where we use the notations of section \ref{sec:state}. We can now compare
the integrand in \eqref{prob1} (with these replacements) with the formula
\eqref{PEE} and we see that it is equal to 
$P_E({\bf x}^{(1)},{\bf x}^{(2)},{\bf x}^{(3)}={\bf x}^{(1)\prime})$ for $m=3$
with the replacement $\tau_2-\tau_1 \to i (t_1-t_2)$ and 
$\tau_2-\tau_3 \to i (t_2-t_1)$. %{\blue What is the meaning of this notation $\prime$ here?}. 
Following the discussion in the
section \ref{sec:state} we know that this quantity can be written
as a determinant, see Eq. \eqref{EM}, thanks to the Eynard-Mehta theorem.
This shows that the EM theorem can also be a useful tool to analyze the
probability of outcomes of quantum measurements in real time for $N$
non-interacting fermions.

\section{Conclusion} \label{sec:conclusion}

In conclusion, we have extended our previous works on the statics of one-dimensional trapped non-interacting fermions to the equilibrium dynamics. In imaginary time we have established an exact mapping between the quantum propagator of the fermions, and the transition probability of a set of $N$ non-crossing classical Ornstein-Uhlenbeck processes. For fermions at thermal equilibrium at temperature $T$, we have shown that there exists a similar mapping, but the corresponding non-crossing Ornstein-Uhlenbeck processes are now periodic in time of period $\beta=1/T$. We have unveiled an extended determinantal structure  of the space time correlations in these two problems, based on an application of the Eynard-Mehta theorem of random matrix theory. Thanks to these properties, we were able to derive the precise universal form of the space-time correlation kernel at the edge of the trapped Fermi gas, at finite temperature. This is a non-trivial extension of the zero temperature result known previously. This also allowed us to introduce and study a new classical stochastic process, which we called the time-periodic Airy$_2$ process that describes the temporal evolution of the top path of $N$ non-intersecting OU processes constrained to return to their initial position after a time period $b \sim 1/T$. The zero temperature version of this process, that is $b \to +\infty$, corresponds to the standard Airy$_2$ process known in the literature. We showed that the joint cumulative distribution function of the process at finite $b$ can be expressed in terms of Fredholm determinants involving the extended edge kernel which we computed explicitly. Although our results were derived for the harmonic trapping potential, we argued that this extended edge kernel, as well as the associated periodic Airy$_2$ process are universal, i.e independent of the details of the trapping potential. This implies the corresponding universality for the classical non-intersecting diffusions in more general potentials.

The special case of the harmonic trapping potential has some additional interesting properties. For example
in the fermion problem, the position and momentum variables play a symmetric role. Hence all the results obtained here for position variables, also hold for the momentum variables. For example we have shown that the (imaginary) time evolution of the fermion positions are isomorphic to the time evolution of the eigenvalues of a Hermitian matrix under the Dyson's Brownian motion. Hence, replacing all position variables by the momentum variables, and using the symmetry between position and momentum, it follows that the (imaginary) time evolution of the fermion momenta is also isomorphic to the time evolution of the eigenvalues of a GUE random matrix under the Dyson's Brownian motion.

As another application of our results, we have studied the real time quantum dynamics of the trapped fermions at the edge of the trapped Fermi gas. We have derived the scaling function of the dynamical density-density correlation, and obtained its 
large time asymptotic decay, found to be a power law $t^{-3/2}$ at $T=0$ and exponential at finite $T$.

The present work raises several interesting questions for future works. For instance, we can ask
whether the periodic Airy$_2$ process introduced here will also appear in other problems of
statistical physics or probability theory. This may not be too far-fetched given the fact that the 
standard Airy$_2$ process itself appears in many different contexts, such as in the large time limit of the (1+1)-dimensional KPZ growth models. In fact our previous results provide a tantalizing hint in that direction. Indeed in \cite{DLMS15} we showed that 
the PDF of the time-periodic Airy$_2$ process at a single point, ${\cal A}^b_2(0)$, coincides exactly (up to a convolution with a Gumbel distribution) with the height of the KPZ equation at time $t_{KPZ}$ with droplet initial condition, with the
correspondence $t_{\rm KPZ}=b^3$. Whether these facts can be tied together is an open subject for future research.

\acknowledgements

We thank D. Bernard, A. Borodin and D. S. Dean for useful discussions.

%\newpage

\begin{appendix}

\section{Non-crossing OU paths, Dyson's Brownian motion and trapped fermions in imaginary time}
\label{app:DBM}

In this appendix we first recall the relation between $N$ non-crossing OU paths and
the Dyson's Brownian motion. Then we establish a relation between the DBM and the quantum imaginary time
propagator of $N$ non-interacting fermions in a harmonic trap. 

%We define the
%transition probability $\hat P^{(N)}_{\rm OU}(x_1,\cdots,x_N;\tau| y_1,\cdots,y_N;0)$ that a set of
%$N$ distinguishable OU processes $x_i(t)$, $i=1,..N$, $0 \leq t \leq \tau$, starting at the initial positions 
%$y_1>y_2>\cdots >y_N$ at time $0$ arrive at $x_1 > x_2 > \cdots >x_N$ 
%at time $\tau$ {\it and} have not crossed each other in the time interval $t \in [0,\tau]$. For $N$ particles the
%generalization of Eq. \eqref{POU3} can be written (in the ordered sector)
%\bea
% \hat P^{(N)}_{\rm OU}(x_1,\cdots,x_N;\tau| y_1,\cdots,y_N;0) & = & e^{- \frac{\mu_0}{2} \sum_{i=1}^N x_i^2}
%\langle x_1,\cdots , x_N | e^{- {\cal H}_N \tau} | y_1, \cdots , y_N \rangle
%e^{ \frac{\mu_0}{2} \sum_{i=1}^N y_i^2} \label{PN1} \\
%&=& e^{- \frac{\mu_0}{2} \sum_{i=1}^N x_i^2} \sum_E \psi_E(x_1,\cdots , x_N)
%\psi_E^*(y_1, \cdots , y_N) e^{- E \tau} e^{ \frac{\mu_0}{2} \sum_{i=1}^N y_i^2} \label{XX}
%\eea
%where the $\psi_E$ are the many-body eigenstates defined in \eqref{slater}. Using the
%determinantal form of $\psi_E$ this transition probability can also be written as
%\be
%\hat P^{(N)}_{\rm OU}(x_1,\cdots,x_N;\tau| y_1,\cdots,y_N;0) = \det P_{\rm OU}(x_i , \tau | y_j, 0) \label{propOU2}
%\ee
%Note that Eqs. \eqref{PN1}, \eqref{propOU2} are valid only in the ordered sector of coordinates,
%and furthermore that the transition probability $\hat P^{(N)}_{\rm OU}$, when integrated over the ordered sector of $x_i$
%coordinates, gives the probability that the $N$ paths do not cross each other up to time $\tau$. 
%
It is interesting to note that the transition probability $P^{(N)}_{\rm OU}$ of the non-crossing OU processes, defined in \eqref{PN1},
can be related to the propagator of the Dyson's Brownian motion (DBM) for the GUE. We recall that the DBM corresponds to considering a complex Hermitian $N \times N$ matrix whose independent entries (both real and imaginary part) perform an OU process in time $\tau$. One can then show the $N$ real eigenvalues $x_i$'s evolve with $\tau$ via the Langevin equation \cite{Dys62,meh91}
\bea
\frac{d x_i(\tau)}{d \tau} = - \mu_0 x_i(\tau) + \sum_{j \neq i} \frac{1}{x_i(\tau) - x_j(\tau)} + \eta_i(\tau) \label{dbm1} \, ,
\eea
where the $\eta_i(\tau)$'s are independent unit white noises. 
If the initial $x_i$'s are ordered, they keep the same order at all times automatically, i.e. they do not cross each other. 
Let us define the probability density
$P^{(N)}_{\rm DBM}(x_1,\cdots,x_N;\tau| y_1,\cdots,y_N;0)$ that the set of eigenvalues,
starting initially from $y_1>y_2>\cdots >y_N$ end up at $x_1 > x_2 > \cdots >x_N$
at time $\tau$ (which is normalized to unity if integrated over the $x_i$ in this ordered sector).
It is then easy to check that the DBM and the non-crossing OU process are
related by (see for example Refs. \cite{Kat15,SchehrRambeau,Baik_book})
\be
P^{(N)}_{\rm DBM}(x_1,\cdots,x_N;\tau| y_1,\cdots,y_N;0) = \frac{ \prod_{1 \leq i <j \leq N} (x_i-x_j)}{\prod_{1 \leq i <j \leq N} (y_i-y_j)} P^{(N)}_{\rm OU}(x_1,\cdots,x_N;\tau| y_1,\cdots,y_N;0) \;. \label{corresp} 
\ee
To check that $P^{(N)}_{\rm DBM}$ is indeed normalized to unity, when integrated over the ordered
sector of the $x_i$'s, we proceed as follows: we insert the exact form for $P^{(N)}_{\rm OU}$ from \eqref{PN1}, \eqref{Satya}
in \eqref{corresp}. Then we note that the many-body ground state wave function is 
\be
\psi_0(x_1,\cdots,x_N)= A_N \prod_{1 \leq i <j \leq N} (x_i-x_j) e^{- \frac{\mu_0}{2} \sum_{i=1}^N x_i^2} \;, \label{psi0}
\ee
where $A_N$ is a normalization constant. 
Using the antisymmetry of the Slater determinants, we can extend the integration over the full space of the $x_i$'s and then use orthonormality of the
eigenstates, leading to the expected normalization to unity.

In addition, using \eqref{corresp} and \eqref{PN1}, we can also relate the DBM directly to the imaginary time evolution of fermions, as seen explicitly in the following relation [using \eqref{psi0}]
\be
P^{(N)}_{\rm DBM}(x_1,\cdots,x_N;\tau| y_1,\cdots,y_N;0) = \frac{ \psi_0(x_1, \cdots, x_N)}{\psi_0(y_1, \cdots, y_N)}
\langle x_1,\cdots , x_N | e^{- {\cal H}_N \tau} | y_1, \cdots , y_N \rangle \;.
  \label{corresp2} 
\ee
This is a direct relation between the classical propagator of the eigenvalues of a complex Hermitian matrix, starting from
arbitrary matrix entries, and the quantum propagator for the positions of the fermions. Note that in this relation the (imaginary) time dependent {\it positions} of the fermions are put in correspondence to the {\it eigenvalues} of the time dependent matrix evolving via the DBM.

%As in the single-particle case in \eqref{PTx2}, one can also establish a connection between this many-body
%quantum JPDF in \eqref{PTxNLast} to the propagator of an $N$-particle OU process defined as follows.
%Indeed, consider $N$ non-intersecting OU processes $x_i(\tau)$, $i=1,..N$, going from the initial positions 
%$y_1>\cdots > y_N$ at time $\tau_0$,
%to the final positions $x_1>\cdots > x_N$, at time $\tau$. The transition probability (the propagator)
%is given by the Karlin-MacGregor formula as a determinant
%\bea
%P^{(N)}_{\rm OU}(x_1,\cdots,x_N;\tau| y_1,\cdots,y_N,\tau_0) = \det P_{\rm OU}(x_i , \tau | y_j, \tau_0) 
%\eea 
%in terms of the single particle OU propagator, given in \eqref{OUprop}. 

\section{More on the time-periodic OU process}
\label{app:more} 

In this section we provide a prescription to construct a time-periodic
OU process. Because of the periodicity constraint in the time direction
it is not evident that this is a stochastic process evolving via a Markov
rule local in time. We restrict here to a single particle for simplicity,
though it can be generalized to $N$-particle system.

\subsection{Langevin equation local in time} 

Consider a single Brownian particle moving in a harmonic potential in one dimension
following the Langevin equation
\eqref{OU1}. If we now impose a constraint that this particle has to return to its initial
position after a fixed time $\beta$ (the OU bridge), can one write an effective Langevin equation, local in time, such that this condition is automatically satisfied by the evolution and that the process is generated with the correct statistical weight. This is indeed possible, as was shown in
Ref. \cite{MO15}, for a large class of constrained stochastic problems. In the case of the time-periodic OU process,
the appropriate Langevin equation reads (see Eq. (26) in Ref. \cite{MO15})
\bea
\frac{d x}{d \tau} = \mu_0 \frac{x_0 - x \cosh( (\beta-\tau) \mu_0 )}{\sinh( (\beta-\tau) \mu_0)} + \eta(\tau) \quad , \quad x(0)=x_0 \;,
\eea 
where $\eta(\tau)$ is the unit white noise. This Langevin equation is such that if we run it with initial condition $x(0)=x_0$ 
we obtain the process conditionned to return in $x_0$ at time $\beta$. In addition to match with the one-time marginal distribution of the time-periodic OU process, we must choose the
initial condition $x_0$ with the measure $P_T(x_0)$ given in Eq. \eqref{statbeta}.
This automatically generates the correct propagator for the time-periodic OU process.

\subsection{Identification as the stationary measure of a confined Edwards-Wilkinson model}

Another method to generate the measure of the time-periodic OU process is as follows.
Consider a single OU process evolving as \eqref{OU1}. The measure over the path $x(\tau)$ can be written
as 
\be
{\cal P}[x(\tau)]  \sim e^{ - \frac{1}{2} \int_0^{\beta} d \tau \left[ (\frac{d x(\tau)}{d \tau} + \mu_0 x(\tau))^2 \right]  } 
\sim e^{ - \frac{1}{2} \int_0^{\beta} d \tau \left[ (\frac{d x(\tau)}{d \tau})^2 + \mu_0^2 x(\tau)^2  \right] } \label{b2} \;.
\ee 
To obtain the second relation we have performed an integration by part and used the periodicity of the path $x(\beta)=x(0)$. One way to generate the measure over paths given in \eqref{b2} is the following. We think of the path
$x(\tau)$ as a fluctuating field in ``space'' $\tau$ with periodic boundary condition of period $\beta$. We now introduce an additional "time evolution" of a field $x(\tau,s)$, with periodicity $\beta$ in the $\tau$ variable, in a 
fictitious time $s$ with evolution equation
\bea
\partial_s x(\tau,s) = \partial_\tau^2 x(\tau,s) - \mu_0^2 x(\tau,s) + \xi(\tau,s) 
\eea 
where $\xi(\tau,s)$ is a Gaussian white noise with zero mean and correlator 
$\langle \xi(\tau_1,s_1) \xi(\tau_2,s_2) \rangle = 2 \delta(\tau_1-\tau_2) \delta(s_1-s_2)$,
periodic in space $\tau$ with period $\beta$. It is easy to see that in the limit of large ``time'' $s$ the
measure of $x(\tau,s)$ converges to an equilibrium measure given precisely by \eqref{b2}. 

\section{How to express $P_E$ in \eqref{PEE} using the Eynard-Mehta theorem}
\label{app:details} 

In this Appendix we provide some details on the derivation of the specific representation
of the EM theorem as stated in the text in Eq. \eqref{EM}. 
We follow the exposition of the theorem given in \cite{BorodinOlshanski} (section 7.4). 
The correspondence of notations is as follows. The time slice index there $k=1,\cdots,m$ is
denoted $\ell=1,\cdots,m$ here. The label of states $n$ there, is called $k$ here. 
The $\phi_{k,n}(x)$ of \cite{BorodinOlshanski} are the same
orthonormal basis $\phi_k(x)$ as introduced here (which is independent of the index $\ell$ 
since our Hamiltonian
is time independent). Furthermore we have the correspondence between the notations
of \cite{BorodinOlshanski} (on the left hand side below) and ours (right hand side) in section \ref{sec:state} 
\bea
&& k \to \ell \quad , \quad n \to k \\
&& c_{k,k+1;n} \to e^{- \epsilon_k (\tau_{\ell+1}- \tau_\ell)} \\
&& \sigma_k(X) \to \det_{1\leq i,j \leq N} \phi_i(x_j) \\
&& v_{k,k+1}(x,y) \to G(x,\tau_{\ell+1}|y,\tau_\ell) 
\eea
where $G(x,\tau_{\ell+1}|y,\tau_\ell)$ is defined in \eqref{propdec}.

The transition probability of the Markov process defined in \cite{BorodinOlshanski} (in their notations,
and using superscripts for
different times) 
%{\red I have changed the bounds on the product downstairs please recheck}
\bea
{\rm Prob}(X^{(1)},\ldots,X^{(m)}) = \frac{\det \phi_i(x^{(1)}_j) \det v_{1,2}(x^{(1)}_i, x^{(2)}_j) \cdots
\det v_{m-1,m}(x^{(m-1)}_i, x^{m)}_j)  \det \phi_i(x^{(m)}_j) }{{\prod_{k=0}^{N-1} e^{- \epsilon_k (\tau_m-\tau_1)} } } \label{C5}
\eea 
where, here, $\det$ stands for $\det_{1\leq i,j \leq N}$ and in \cite{BorodinOlshanski} all $X^{(\ell)}$ are ordered $x^{(\ell)}_1<\ldots< x^{(\ell)}_m$
(here they are denoted ${\bf x}^{(\ell)}$ and are not necessarily ordered). 
Hence one can check that our quantum probability \eqref{PEE}, in one given sector ordered in
all the $X^{(\ell)}$ reads
\bea
P_E({\bf x}^{(1)},\cdots, {\bf x}^{(m)})  = \frac{1}{N!^m} {\rm Prob}(X^{(1)},\cdots, X^{(m)})
\eea 
which is consistent with the normalizations (unordered one here, ordered one in \cite{BorodinOlshanski}). 
{Note that this correspondence, as stated above, gives directly the result when $|E \rangle$ is the ground state.
However it is easily extendable to any $N$-fermion eigenstate $|E \rangle$ of ${\cal H}_N$ parameterized
by a set of occupation numbers $n_k=0,1$, such that $\sum_k n_k=N$. Indeed, one can
just relabel the single particle states $\phi_{a,b}(x)$ of \cite{BorodinOlshanski} so that the
$N$ occupied ones (with $n_k=1$) correspond to the labels $b=0,\cdots,N-1$ (and in that case the denominator in Eq.~(\ref{C5}) changes to $\prod_{k=0}^\infty e^{-n_k \epsilon_k (\tau_m-\tau_1)}$). 
This leads to the statement of the EM theorem in terms of occupation numbers as
given in Eq. (\ref{EM}) in the text.} {Finally note that
\cite{BorodinOlshanski} focuses for simplicity on position variables $x_j^{(\ell)}$ on a discrete lattice,
but is straightforward to extend it to continuous space $x_j^{(\ell)} \in \mathbb{R}$ and considering probability densities instead of probability at a discrete site.}
%, given that we have checked all normalization conditions.}

\section{Self-reproducibility of the extended kernel} 
\label{app:repro} 

It is interesting to note the following self-reproducibility property of the extended kernel
defined in Eqs. \eqref{Kext10}-\eqref{Kext20}. 
Using the orthonormality of the
wave functions, and the fact that $n_k^2=n_k$ for fermion occupation numbers, we find 
\bea
&& \int dy \,  K(x,\tau;y, \tau';\{ n_k \}) K(y,\tau'; z,\tau'';\{ n_k \}) = \eta_{\tau,\tau',\tau''} K(x,\tau; z,\tau'';\{ n_k \}) \eea 
with 
\bea
&& \eta_{\tau \geq \tau' \geq \tau''} = + 1 \quad , \quad \eta_{\tau'' > \tau' > \tau} = - 1
\eea 
and $\eta_{\tau,\tau',\tau''}=0$ in all other cases. 

\section{Determinantal form of general correlation functions} 
\label{app:det}

In the text we have defined multi-time correlations involving a single particle
at each time slice, i.e. the spatio-temporal correlation of the fermion density, in Eqs. \eqref{Rm1}-\eqref{Rm2}.
Here we define more general correlation functions, involving arbitrary number $\ell_j$ of
fermions in each time slice $\tau_j$, $j=1,\cdots, m$, and give a determinantal expression for such a correlation.
For $\tau_1 < \cdots < \tau_m$ one defines the correlation function in a fixed eigenstate $|E \rangle$ of the $N$-body Hamiltonian, labeled by a set of occupation numbers $\{ n_k \}$ 
\bea
&& \hat R_{\ell_1,..\ell_m}(\{ x^{(1)}_1, \tau_1; \cdots ; x^{(1)}_{\ell_1}, \tau_1\}, \cdots , 
\{ x^{(m)}_1, \tau_m, \cdots ,x^{(m)}_{\ell_m}, \tau_m\}) 
\\
&& : = {\prod_{k=1}^m \frac{N!}{(N-n_{\ell_k})!} }
 \int \prod_{k=1}^m {\prod_{i=\ell_k+1}^{N} } dx_i^{(k)} P_E({\bf x}^{(1)},\cdots, {\bf x}^{(m)})
\eea 
where $P_E$ is given in \eqref{PEE}.
The Eynard-Mehta theorem implies that it is given by
\bea
&& {\hat R_{\ell_1,\cdots,\ell_m}(\{ x^{(1)}_1, \tau_1; \cdots ;x^{(1)}_{\ell_1}, \tau_1\}, \cdots , 
\{ x^{(m)}_1, \tau_1; \cdots ;x^{(m)}_{\ell_m}, \tau_m\}) }\\
&&=  \det_{ 1 \leq p, q \leq m, 1 \leq i \leq \ell_p, 1 \leq j \leq \ell_q }
K( x^{(p)}_i , \tau_p ; x^{(q)}_j , \tau_q ;\{ n_k \})) \;.
\eea 

%
%
%\section{Rewriting of the kernel}
%

%{\red what is below I suspect is not correct, but it would be nice to have a correct
%version} 
%One can now introduce the generalized weight
%\bea
%\Sigma_{b,u}(v) &=& \sigma_b(v) e^{- (u_i-u_j)} \quad , \quad u_i \geq u_j \\
%&=& (\sigma_b(v) -1) e^{- (u_i-u_j)} \quad , \quad u_i < u_j \\
%\sigma_b(v) &=& \frac{1}{e^{-b v}+1} 
%\eea 
%Then, as can be checked by expanding in traces and performing integrals
%over the $s,s'$ variables, one can write an equivalent formula
%\bea
%Prob( {\cal A}^b_{2}(u_1) < s_1, {\cal A}^b_{2}(u_2) < s_2) 
%&=& {\rm Det} \left[ \delta_{ij} I - \bar K_{u_i-u_j;s_i,s_j}  \right]_{1 \leq i,j \leq 2}
%\eea 
%in terms of the matrix kernel
%\bea
%\bar K_{u_i-u_j;s_i,s_j}(v,v') = K_\Ai(s_i + v, s_j + v') \Sigma_{b,u_i-u_j}(v) \quad , \quad 1 \leq i,j \leq 2
%\eea 
%which involves the Airy kernel and the generalized weight. \\
%

\end{appendix}


\begin{thebibliography}{6}
\bibitem{PraSpo02}
M. Pr\"ahofer, H. Spohn, {\it Scale Invariance of the PNG Droplet and the Airy Process}, J. Stat. Phys.,
{\bf 108}, 1071 (2002).
\bibitem{Joh03}
K. Johansson, {\it Discrete polynuclear growth and determinantal processes}, Comm. Math. Phys. {\bf 242}, 277
(2003).


%\bibitem{AGZ07}
%G. W. Anderson, A. Guionnet, and O. Zeitouni. An introduction to random
%matrices. Vol. 118. Cambridge Studies in Advanced Mathematics. Cambridge:
%Cambridge University Press, 2010, pp. xiv+492.

\bibitem{Dys62}
F. J. Dyson, {\it A Brownian-motion model for the eigenvalues of a random matrix}, J. Math. Phys. {\bf 3}, 1191 (1962).


\bibitem{TW07}
C. Tracy, H. Widom, {\it Nonintersecting Brownian excursions}, Ann. Appl. Probab. {\bf 17}, 953 (2007).

\bibitem{CH11}
I. Corwin, A. Hammond, {\it Brownian Gibbs property for Airy line ensembles}, {\it Invent. Math.} {\bf 195}, 441 (2014).  

\bibitem{spohn_oup}
P. L. Ferrari, H. Spohn, {\it Random growth models}, in {\it The Oxford Handbook of Random Matrix Theory}, edited by G. Akemann, J. Baik, Ph. di Francesco, (Oxford University Press, Oxford, 2011). 

\bibitem{Joh05}
K. Johansson, {\it The arctic circle boundary and the Airy process}, Ann. Probab. {\bf 33}, 1 (2005).

\bibitem{Ferrari}
P. L. Ferrari, {\it The universal Airy$_1$ and Airy$_2$ processes in the Totally Asymmetric Simple Exclusion Process}, in Proceedings {\it Integrable Systems and Random Matrices: In Honor of Percy Deift}, Contemporary Mathematics {\bf 458}, 321 (2008).  

\bibitem{CH16}
I. Corwin, A. Hammond, {\it KPZ line ensembles}, arXiv:1312.2600, to appear in Probab. Their. and Related Fields.


\bibitem{QR14}
J. Quastel, D. Remenik, {\it Airy processes and variational problems}. In
{\it Topics in Percolative and Disordered Systems } Springer Proc. Math. Stat. {\bf 69} 121Ð171. Springer, New York (2014). 

\bibitem{Baik_book}
J. Baik, P. Deift, T. Suidan, {\it Combinatorics and Random Matrix theory}, AMS {\bf 172}, (2016). 

\bibitem{Risken}
H. Risken, {\it The Fokker-Planck equation}, Springer series in synergetics, vol. 18 (Springer, Berlin, NY) 1984.

\bibitem{footnote0} 
{Note that the bridge condition is not strictly necessary here. What is needed is to
consider a very long OU path (infinite from $]-\infty,+\infty[$ in the limit) and
ask about the PDF that it passes by $x$. In that limit, any (reasonable) boundary condition
will produce the same stationary measure \eqref{Gibbs1}. The same statement holds for the
$N$ non-crossing path problem, see below.}

\bibitem{KM59}
S. Karlin, J. McGregor, {\it Coincidence probabilities}, Pacific J. Math. {\bf 9}, 1141 (1959).

\bibitem{bray_winkler}
A. J. Bray, K. Winkler, {\it Vicious walkers in a potential}, J. Phys. A: Math. Gen. {\bf 37}, 5493 (2004). 



%\bibitem{Fisher84}
%M.~ E. Fisher, \emph{Walks, walls, wetting, and melting}, J. Stat. Phys. 
%\textbf{34}, 667 (1984).

\bibitem{meh91}{M.~L. Mehta, {\it Random Matrices} 
(Academic Press, Boston, 1991).}



\bibitem{watermelon_us}
G. Schehr, S. N. Majumdar, A. Comtet, J. Randon-Furling, {\it Exact distribution of the maximal height of p vicious walkers}, Phys. Rev. Lett. {\bf 101}, 150601 (2008).

\bibitem{KIK08}
N. Kobayashi, M. Izumi, M. Katori, {\it Maximum distributions of bridges of non-colliding Brownian paths}, Phys. Rev. E {\bf 78}, 051102 (2008). 

\bibitem{FMS11}
P. J. Forrester, S. N. Majumdar, G. Schehr, {\it Non-intersecting Brownian walkers and Yang-Mills theory on the sphere}, Nucl.Phys. B {\bf 844}, 500 (2011). 


\bibitem{SMCF13}
G. Schehr, S. N. Majumdar, A. Comtet, P. J. Forrester, Reunion probability of $N$ vicious walkers: typical and large fluctuations for large $N$,  J. Stat. Phys. {\bf 150}, 491 (2013). 


\bibitem{Cor12}
I. Corwin, {\it The Kardar-Parisi-Zhang equation and
universality class}, Rand. Mat.: Theo. Appl. {\bf 1}, 1130001 (2012).



%\bibitem{Sch2012}
%G. Schehr, Extremes of N vicious walkers for large $N$: application to the directed polymer and KPZ interfaces, J. %Stat. Phys. 149(3), 385-410 (2012). 

\bibitem{KK10}
T. Kriecherbauer, J. Krug, {\it A pedestrian's view on interacting particle systems, KPZ universality, and random matrices}, J. Phys. A: Math. Theor. {\bf 43}, 403001 (2010).

\bibitem{TW94}
C. A. Tracy, H. Widom, {\it Level-spacing distributions and the Airy kernel}, Commun. Math. Phys., {\bf 159}, 151 (1994).

\bibitem{AM04}
M. Adler, P. van Moerbeke, {\it PDEs for the joint distributions of the Dyson, Airy and Sine processes}, Ann. Probab. {\bf 33}, 1326 (2004). 

\bibitem{MMSV14} R. Marino, S. N. Majumdar, G. Schehr, P. Vivo, {\it Phase transitions and edge scaling of number variance in Gaussian random matrices}, Phys. Rev. Lett. {\bf 112}, 254101 (2014).

\bibitem{MMSV16}
R. Marino, S. N. Majumdar, G. Schehr, P. Vivo, {\it Number statistics for $\beta$-ensembles of random matrices: applications to trapped fermions at zero temperature}, Phys. Rev. E {\bf 94}, 032115 (2016).

\bibitem{DLMS15} D.~S. Dean, P. Le Doussal, S.~N. Majumdar, G. Schehr, {\it Finite temperature free fermions and the Kardar-Parisi-Zhang equation at finite time}, Phys. Rev. Lett. {\bf 114}, 110402 (2015).

\bibitem{dea15b} D.~S. Dean, P. Le Doussal, S.~N. Majumdar, G. Schehr, {\it Universal ground state properties of free fermions in a $d$-dimensional trap}, Europhys. Lett. {\bf 112}, 60001 (2015).

\bibitem{DLMS16} D.~S. Dean, P. Le Doussal, S.~N. Majumdar, G. Schehr, {\it Non-interacting fermions at finite temperature in a d-dimensional trap: universal correlations}, arXiv:1609.04366, Phys. Rev. A. {\bf 94} 063622 (2016). 

\bibitem{dea16}
D.~S. Dean, P. Le Doussal, S.~N. Majumdar, G. Schehr, {\it Statistics of the maximal distance and momentum in a trapped Fermi gas at low temperature}, preprint arXiv:1612.03954.

\bibitem{Eis13}
V. Eisler, {\it Universality in the full counting statistics of trapped fermions}, Phys. Rev. Lett. {\bf 111}, 080402 (2013).

\bibitem{Castillo14}
%Spectral order statistics of Gaussian random matrices: large deviations 
%for trapped fermions and associated phase transitions
I. P. Castillo, {\it Spectral order statistics of Gaussian random matrices: large deviations for trapped fermions and associated phase transitions} Phys. Rev. E {\bf 90}, 040102 (2014).


\bibitem{FW71}
A. L. Fetter, J. D. Walecka, {\it Quantum Theory of Many-Particle Systems}, (McGraw-Hill, 1971)

\bibitem{GiamBook}
T. Giamarchi, {\it Quantum Physics in One Dimension}, Oxford Clarendon Press (2004).

\bibitem{Pereira}
R.G. Pereira, {\it Long time correlations of nonlinear Luttinger liquid}, Int. J. Mod. Phys. B {\bf 26}, 1244008 (2012) 
% Long time correlations of nonlinear Luttinger liquid

\bibitem{Xavier}
%Boundary versus bulk behavior of time-dependent correlation functions in one-dimensional quantum systems
I. S. Eliëns, F. B. Ramos, J. C. Xavier, and R. G. Pereira, {\it Boundary versus bulk behavior of time-dependent correlation functions in one-dimensional quantum systems}, Phys. Rev. B {\bf 93}, 195129 (2016).

\bibitem{Sims} 
%Decay of Determinantal and Pfaffian Correlation Functionals in One-dimensional Lattices
R. Sims and S. Warzel, {\it Decay of Determinantal and Pfaffian Correlation Functionals in One-dimensional Lattices}, Commun. Math. Phys. (347), 903 (2016). 

\bibitem{Stolze}
J. Stolze, A. Noppert, G. Muller, {\it Gaussian, exponential, and power-law decay of time-dependent correlation functions in quantum spin chains}, Phys. Rev. B {\bf 52}, 4319 (1995).
%Gaussian, exponential, and power-law decay of time-dependent correlation functions in quantum spin chains.

\bibitem{Mac94}
A. M. S. Macedo, {\it Universal parametric correlations at the soft edge of the spectrum of random matrix ensembles},  Europhys. Lett. {\bf 26}, 641 (1994).

\bibitem{Borodin1}
A. Borodin, {\it Determinantal point processes}, in {\it The Oxford Handbook of Random Matrix Theory}, 
G. Akemann, J. Baik, P. Di Francesco (Eds.), Oxford University Press, Oxford (2011).



\bibitem{EM98}
B. Eynard, M. L. Mehta, {\it Matrices coupled in a chain. I. Eigenvalue correlations}, J.
Phys. A: Math. Gen. {\bf 31}, 4449 (1998).


\bibitem{footnotenew} 
{Note that here in the quantum propagator the final point is on the left, and the initial on the right, 
which is opposite to the notations we used in \cite{DLMS16}.} 



\bibitem{nad09} C. Nadal, S.~N. Majumdar, {\it Non-intersecting Brownian Interfaces and Wishart Random Matrices}, Phys. Rev. E {\bf 79}, 061117 (2009).





\bibitem{BorodinOlshanski}
A. Borodin, G. Olshanski, {\it Markov processes on partitions}, Probab. Theory Rel. {\bf 135}, 84 (2006).

\bibitem{Johansson2005}
K. Johansson, {\it Random matrices and determinantal processes}, in Lecture Notes of the Les Houches Summer School 2005 (A. Bovier, F. Dunlop, A. van Enter, F. den Hollander, and J. Dalibard, eds.), Elsevier Science, (2006); arXiv:math-ph/0510038. 


\bibitem{For10}{ P.~J. Forrester, 
{\it Log-Gases and Random Matrices} 
(London Mathematical Society monographs, 2010).  }








\bibitem{Joh07}
{K. Johansson, {\it From Gumbel to Tracy-Widom}, Probab. Theory Rel. {\bf 138}, 75 (2007).}
%From Gumbel to Tracy-Widom

%\bibitem{Dyson}
%F.J. Dyson, J. Math. Phys. {\bf 3} 1191 (1962); ibid 1198.


 
\bibitem{gaudin}
M. Gaudin, {\it Une d\'emonstration simplifi\'ee du th\'eor\`eme de Wick en m\'ecanique quantique}, Nucl. Phys. {\bf 15}, 89 (1960).

\bibitem{Kat15}
M. Katori, {\it Bessel processes, Schramm-Loewner evolution, and the Dyson model}, Vol. 11 of Springer Briefs in Mathematical Physics, Springer, (Singapore), (2015).


\bibitem{SchehrRambeau}
J. Rambeau, G. Schehr, {\it Distribution of time at which $N$ vicious walkers reach their maximal height},
Phys. Rev. E {\bf 83} 061146 (2011). 




\bibitem{MO15}
S. N. Majumdar, H. Orland, {\it Effective Langevin equations for constrained stochastic processes}, J. Stat. Mech. P06039 (2015). 







%\bibitem{footnote1}
%Note that $F_2^b(s)$ here is called $Q_b(s)$ in Eq. (137) of \cite{DLMS16} 
%and the scaled edge density, ${\cal K}_b^{\rm edge}(s,s)$, is denoted by $F_{1,b}(s)$ 
%in Eq. (126) of \cite{DLMS16}




%%%% a verifier si ref below necessaires 

 \end{thebibliography}
 \end{document}